\documentclass[useAMS,usenatbib]{mn2e}
\usepackage{graphicx,txfonts,amssymb,amsfonts,natbib,times,hyperref}

\begin{document}

\title[Constraints on primordial non-Gaussianity from future surveys]
  {The clustering of galaxies and galaxy clusters: constraints on primordial non-Gaussianity from future wide-field surveys}

\author[C. Fedeli et al.]{C. Fedeli$^{1,2}$, C. Carbone $^{1,3}$, L. Moscardini$^{1,3}$, and A. Cimatti$^1$\\
$^1$ Dipartimento di Astronomia, Universit\`a di Bologna, Via Ranzani 1, I-40127 Bologna, Italy\\
$^2$ Department of Astronomy, University of Florida, 312 Bryant Space Science Center, Gainesville, FL 32611\\      
$^3$  INFN, Sezione di Bologna, Viale Berti Pichat 6/2, I-40127 Bologna, Italy\\
\\
cosimo.fedeli@astro.ufl.edu, carmelita.carbone@unibo.it, lauro.moscardini@unibo.it, a.cimatti@unibo.it\\}

\maketitle

\begin{abstract}
We investigate the constraints on primordial non-Gaussianity with varied bispectrum shapes that can be derived from the power spectrum of galaxies and clusters of galaxies detected in future wide field optical/near-infrared surveys. Having in mind the proposed ESA space mission \emph{Euclid} as a specific example, we combine the spatial distribution of spectroscopically selected galaxies with that of weak lensing selected clusters. We use the physically motivated halo model in order to represent the correlation function of arbitrary tracers of the Large Scale Structure in the Universe. As naively expected, we find that galaxies are much more effective in jointly constrain the level of primordial non-Gaussianity $f_\mathrm{NL}$ and the amplitude of the matter power spectrum $\sigma_8$ than clusters of galaxies, due to the much lower abundance of the latter that is not adequately compensated by the larger effect on the power spectrum. Nevertheless, combination of the galaxy power spectrum with the cluster-galaxy cross spectrum can decrease the error on the determination of $f_\mathrm{NL}$ by up to a factor of $\sim 2$. This decrement is particularly evident for the less studied non-Gaussian bispectrum shapes, the so-called enfolded and the orthogonal ones. Setting constraints on these models can shed new light on various aspects of the physics of the early Universe, and it is hence of extreme importance. By combining the power spectra of clusters and galaxies with the cluster-galaxy cross spectrum we find constraints on primordial non-Gaussianity of the order $\Delta f_\mathrm{NL} \sim $ a few, competitive and possibly superior to future CMB experiments.
\end{abstract}

\section{Introduction}\label{sct:introduction}

It is now commonly accepted that the formation of structures in the Universe originated from seed density fluctuations in the dark matter fluid that were laid down during inflation \citep*{HA83.1}. Gravitational instability has the effect of amplifying density perturbations, that at a given point enter the non-linear regime and collapse to form clumps, voids, and more in general the complex Large Scale Structure (LSS henceforth) that we observe today. For a given distribution of the initial conditions, the statistical properties of the LSS are determined uniquely by the subsequent expansion history of the Universe that is, ultimately, on its matter and energy content. In the past decade much effort has been directed toward understanding the effect of dark energy on the formation of structures (see \citealt*{CU09.1,GR09.2,SA09.1}; \citealt{DE10.2}; \citealt*{MO10.1} for recent examples) in order to gain insights on its nature, specifically whether it truly is a cosmological constant or it does have some kind of evolution with cosmic time. 

After several pioneering works \citep{ME90.1,MO91.1,WE92.1}, only recently has the question of the initial conditions gained renewed attention, specifically about the shape of the primordial density fluctuations distribution. While the simplest models of inflation (single slow-rolling scalar field) predicts this distribution to be virtually indistinguishable from a Gaussian, a plethora of more complex models have been proposed that, in addition to solving the standard cosmological problems, allow for significant and possibly scale dependent deviations from Gaussianity. The study of non-Gaussian cosmologies and the effect they have on the formation and evolution of cosmic structures is thus extremely important in order to rule out inflationary models, and hence to have a better handle on the physics of the early Universe. Moreover, studies on possible detectability of primordial non-Gaussianity are very timely, given the recent claims for a positive skewness of the primordial density fluctuations distribution coming from the Cosmic Microwave Background (CMB) and from the angular correlation function of radio-selected quasars \citep{XI10.1}.

The problem of constraining deviations from primordial Gaussianity by means different from the CMB intrinsic anisotropies has recently attracted much efforts in the literature, with studies directed towards the abundance of non-linear structures (\citealt*{MA00.2}; \citealt{VE00.1}; \citealt*{MA04.1}; \citealt{GR07.1,GR09.1,MA10.2}), halo biasing (\citealt{DA08.1,MC08.1}; \citealt*{FE09.1}), galaxy bispectrum \citep{SE07.2,JE09.1}, mass density distribution \citep{GR08.2} and topology \citep{MA03.2,HI08.2}, cosmic shear (\citealt*{FE10.1}, Pace et al. 2010), integrated Sachs-Wolfe effect \citep*{AF08.1,CA08.1}, Ly$\alpha$ flux from low-density intergalactic medium \citep{VI09.1}, $21-$cm fluctuations \citep*{CO06.2,PI07.1} and reionization \citep{CR09.1}.

In this work we focused attention on the spatial distribution of galaxies and galaxy clusters as tracers of the LSS. We were particularly interested in comparing the performances of the power spectra of the two individual tracers in constraining primordial non-Gaussianity, and evaluate the improvements in forecasted constraints given by the addition of the cluster-galaxy cross power spectrum. Throughout the paper we assumed a fiducial future all-sky optical/near-infrared survey on the model of \emph{Euclid} \citep{LA09.1,BE10.1}. In order to fully exploit the potentials of both the imaging and the spectroscopy part of \emph{Euclid} we considered galaxies as spectroscopically selected according to their H$\alpha$ flux, and galaxy clusters as selected to be the high signal-to-noise ratio (S/N) peaks in full-sky cosmic shear maps. This approach has the advantage of allowing treatment of the galaxy and cluster samples as independent. The results obtained here are relevant for other planned missions with a similar concept to \emph{Euclid}, such as the NASA Wide Field Infrared Survey (WFIRST).

The rest of the paper is organized as follows. In Sections \ref{sct:ng} and \ref{sct:mf} we summarize the non-Gaussian models that we have explored in this work, as well as their effect on the mass function and large scale bias of dark matter halos. In Section \ref{sct:halomodel} we describe the halo model, the physically motivated framework that we adopted for modeling the power spectrum of clusters and galaxies as well as the cluster-galaxy cross spectrum. In Section \ref{sct:results} we summarize our results and in Section \ref{sct:discussion} we discuss them, with particular emphasis on alternative survey configurations. Finally, in Section \ref{sct:conclusions} we present our conclusions. Throughout this work we refer to the fiducial cosmological model as the one defined by the parameter set derived by the latest analysis of the WMAP data \citep{KO10.1}. This means that density parameters for matter, cosmological constant, and baryons are equal to $\Omega_{\mathrm{m},0}=0.272$, $\Omega_{\Lambda,0} = 0.728$, and $\Omega_{\mathrm{b},0}=0.046$, respectively, the Hubble constant equals $h\equiv H_0/(100 \;\mathrm{km\;s}^{-1}\;\mathrm{Mpc}^{-1})=0.704$, and the normalization of the matter power spectrum is set by $\sigma_8=0.809$.

\section{Non-Gaussian cosmologies}\label{sct:ng}

Extensions of the most standard model of inflation \citep{ST79.1,GU81.1,LI82.1} can produce substantial deviations from a Gaussian distribution of primordial density and potential fluctuations (see \citealt{BA04.1,CH10.1,DE10.1} for recent reviews). The amount and shape of this deviation depend critically on the kind of non-standard inflationary model that one has in mind, as will be detailed later on.

A particularly convenient (although not unique) way to describe generic deviations from a Gaussian distribution consists in writing the gauge-invariant Bardeeen's potential $\Phi$ as the sum of a Gaussian random field and a quadratic correction \citep{SA90.1,GA94.1,VE00.1,KO01.1}, according to

\begin{equation}\label{eqn:ng}
\Phi = \Phi_\mathrm{G} + f_\mathrm{NL} * \left( \Phi_\mathrm{G}^2 - \langle \Phi_\mathrm{G}^2 \rangle \right).
\end{equation}
The parameter $f_\mathrm{NL}$ in Eq. (\ref{eqn:ng}) determines the amplitude of non-Gaussianity, and it is in general dependent on the scale. The symbol $*$ denotes convolution between functions, and reduces to standard multiplication upon constancy of $f_\mathrm{NL}$. In the following we adopted the large-scale structure convention (as opposed to the CMB convention, see \citealt{AF08.1,CA08.1}; \citealt*{PI09.1} and \citealt{GR09.1}) for defining the fundamental parameter $f_\mathrm{NL}$. According to this, the primordial value of $\Phi$ has to be linearly extrapolated at $z = 0$, and as a consequence the constraints given on $f_\mathrm{NL}$ by the CMB have to be raised by $\sim 30$ per cent to comply with this paper's convention (see also \citealt*{FE09.1} for a concise explanation).

In the case in which $f_\mathrm{NL} \ne 0$ the potential $\Phi$ is a random field with a non-Gaussian probability distribution. Therefore, the field itself cannot be described by the power spectrum $P_\Phi({\bf k}) = Bk^{n-4}$ alone, rather higher-order moments are needed. The dominant higher-order contribution is generically given by the bispectrum $B_\Phi({\bf k}_1,{\bf k}_2,{\bf k}_3)$. Only in very peculiar situations the bispectrum vanishes, and one has to resort to the trispectrum or higher-order correlations. The bispectrum is the Fourier transform of the three-point correlation function $\langle \Phi({\bf k}_1)\Phi({\bf k}_3)\Phi({\bf k}_3) \rangle$ and it can hence be implicitly defined as

\begin{equation}
\langle \Phi({\bf k}_1)\Phi({\bf k}_3)\Phi({\bf k}_3) \rangle \equiv (2\pi)^3\delta_\mathrm{D}\left( {\bf k}_1+{\bf k}_2+{\bf k}_3 \right) B_\Phi({\bf k}_1,{\bf k}_2,{\bf k}_3).
\end{equation}

As mentioned above understanding the shape of non-Gaussianity is of fundamental importance in order to pinpoint the physics of the early universe and the evolution of the inflaton field in particular. For this reason, in this work we considered four different shapes of the potential bispectrum, arising from different modifications of the standard inflationary scenario. They are all described in the following.

\subsubsection*{Local shape}

The standard single-field inflationary scenario generates negligibly small deviations from Gaussianity. These deviations are said to be of the local shape, and the related bispectrum of the Bardeen's potential is maximized for \emph{squeezed} configurations, where one of the three wavevectors has much smaller magnitude than the other two. In this case the parameter $f_\mathrm{NL}$ must be a constant (see however \citealt{BY10.2,BY10.1}), and it is expected to be of the same order of the slow-roll parameters \citep*{FA93.1}, that are very close to zero. 

However non-Gaussianities of the local shape can also be generated in the case in which an additional light scalar field, different from the inflaton, contributes to the observed curvature perturbations \citep*{BA04.2}. This happens, for instance, in curvaton models \citep*{SA06.1,AS07.1} or in multi-fields models \citep*{BA02.2,BE02.1,HU09.1}. In this case the parameter $f_\mathrm{NL}$ is allowed to be substantially different from zero, and the bispectrum of the primordial potential assumes the simple form

\begin{equation}
B_\Phi({\bf k}_1,{\bf k}_2,{\bf k}_3) = 2f_\mathrm{NL} B^2 \left[ k_1^{n-4}k_2^{n-4} + k_1^{n-4}k_3^{n-4} + k_2^{n-4}k_3^{n-4} \right].
\end{equation}
Due to its simplicity, the local model is the most studied one, especially in terms of cosmological simulations (see however \citealt*{WA10.1}).

\subsubsection*{Equilateral shape}

In some inflationary models the kinetic term of the inflaton Lagrangian is not standard, containing higher-order derivatives of the field itself. One significant example of this is the DBI model (\citealt*{AL04.1,SI04.1}, see also \citealt{AR04.1}; \citealt*{SE05.1,LI08.1}). In this case the primordial bispectrum is maximized for configurations where the three wavevectors have approximately the same amplitude, and it takes the form \citep{CR07.1}

\begin{eqnarray}\label{eqn:eqb}
B_\Phi({\bf k}_1,{\bf k}_2,{\bf k}_3) &=&  6f_\mathrm{NL} B^2 \gamma({\bf k}_1,{\bf k}_2,{\bf k}_3) \left[ k_1^{(n-4)/3}k_2^{2(n-4)/3}k_3^{n-4} \right. + 
\nonumber\\
&+& k_3^{(n-4)/3}k_1^{2(n-4)/3}k_2^{n-4} + k_2^{(n-4)/3}k_3^{2(n-4)/3}k_1^{n-4} + 
\nonumber\\
&+& k_1^{(n-4)/3}k_3^{2(n-4)/3}k_2^{n-4} + k_2^{(n-4)/3}k_1^{2(n-4)/3}k_3^{n-4} + 
\nonumber\\
&+& k_3^{(n-4)/3}k_2^{2(n-4)/3}k_1^{n-4} -k_1^{n-4}k_2^{n-4} - k_1^{n-4}k_3^{n-4} -
\nonumber\\
&-& \left. k_2^{n-4}k_3^{n-4} - 2k_1^{2(n-4)/3}k_2^{2(n-4)/3}k_3^{2(n-4)/3}\right]. 
\end{eqnarray}

The function $\gamma({\bf k}_1,{\bf k}_2,{\bf k}_3)$ in the first line of the previous equation represents a \emph{running} of the parameter $f_\mathrm{NL}$ that we have the liberty to insert since this parameter is not forced to be constant in the present case. This running has been considered as an actual part of the equilateral bispectrum in all the calculations that follow. It reads \citep{CH05.2,LO08.1,CR09.1}

\begin{equation}
\gamma({\bf k}_1,{\bf k}_2,{\bf k}_3) = \left( \frac{k_1+k_2+k_3}{k_\mathrm{CMB}} \right)^{-2\kappa}.
\end{equation}
We adopted the exponent $\kappa=-0.2$, that increase the level of non-Gaussianity at scales smaller than that corresponding to $k_\mathrm{CMB} = 0.086 h$ Mpc$^{-1}$. This coincides with the larger multipole used in the CMB analysis by the WMAP team \citep{KO09.1,KO10.1}, $\ell\sim 700$. When referring to the equilateral shape in the rest of this paper we always mean the bispectrum given by Eq. (\ref{eqn:eqb}) with $\kappa=-0.2$, unless otherwise noted.

\subsubsection*{Enfolded shape}

For deviations from Gaussianity evaluated in the regular Bunch-Davies vacuum state, the primordial potential bispectrum is of local or equilateral shape, depending on whether or not higher-order derivatives play a significant role in the evolution of the inflaton field. If the Bunch-Davies vacuum hypothesis is dropped, the resulting bispectrum is maximal for \emph{squashed} configurations \citep{CH07.1,HO08.1}. 

\cite*{ME09.1} found a template that describes very well the properties of this enfolded-shape bispectrum (see also \citealt{CR10.1}), that reads

\begin{eqnarray}
B_\Phi({\bf k}_1,{\bf k}_2,{\bf k}_3) &=& 6f_\mathrm{NL} B^2 \left[ k_1^{n-4}k_2^{n-4} + k_1^{n-4}k_3^{n-4} + k_2^{n-4}k_3^{n-4} + \right.
\nonumber\\
&+& 3 \left( k_1^{n-4}k_2^{n-4}k_3^{n-4} \right)^{2/3}-
\nonumber\\
&-&\left(k_1^{(n-4)/3}k_2^{2(n-4)/3}k_3^{n-4} +  k_3^{(n-4)/3}k_1^{2(n-4)/3}k_2^{n-4} +\right.
\nonumber\\
&+& k_2^{(n-4)/3}k_3^{2(n-4)/3}k_1^{n-4} + k_1^{(n-4)/3}k_3^{2(n-4)/3}k_2^{n-4} 
\nonumber\\
&+& \left.\left. k_2^{(n-4)/3}k_1^{2(n-4)/3}k_3^{n-4} + k_3^{(n-4)/3}k_2^{2(n-4)/3}k_1^{n-4}\right)\right].
\end{eqnarray}
No running of the $f_\mathrm{NL}$ parameter was introduced for both the enfolded and orthogonal shapes. The reasons are discussed further below.

\subsubsection*{Orthogonal shape}

A shape of the bispectrum can be constructed that is nearly \emph{orthogonal} to both the local and equilateral forms \citep*{SE10.1}. In this case the potential bispectrum is well approximated by the following template

\begin{eqnarray}
B_\Phi({\bf k}_1,{\bf k}_2,{\bf k}_3) &=& 6f_\mathrm{NL} B^2 \left[ 3\left(k_1^{(n-4)/3}k_2^{2(n-4)/3}k_3^{n-4} + \right.\right. 
\nonumber\\
&+&k_3^{(n-4)/3}k_1^{2(n-4)/3}k_2^{n-4} + k_2^{(n-4)/3}k_3^{2(n-4)/3}k_1^{n-4}
\nonumber\\
&+& k_1^{(n-4)/3}k_3^{2(n-4)/3}k_2^{n-4} + k_2^{(n-4)/3}k_1^{2(n-4)/3}k_3^{n-4} 
\nonumber\\
&+& \left.\left. k_3^{(n-4)/3}k_2^{2(n-4)/3}k_1^{n-4}\right)-8\left( k_1^{n-4}k_2^{n-4}k_3^{n-4} \right)^{2/3}\right.-
\nonumber\\
&-&\left. 3 \left(k_1^{n-4}k_2^{n-4} + k_1^{n-4}k_3^{n-4} + k_2^{n-4}k_3^{n-4}\right) \right].
\end{eqnarray}
Constraints on the level of non-Gaussianity compatible with the CMB in the local, equilateral and orthogonal scenarios were recently given by the WMAP team \citep{KO10.1}, while constraints on enfolded non-Gaussianity from LSS were given by \citet{VE09.1}\\
\\\indent
Although there is no theoretical prescription against a running of the $f_\mathrm{NL}$ parameter with the scale in the enfolded and orthogonal shapes, we decided not to include one. The reason for this is that there is no first principle that can guide one in the choice of a particular kind of running, and until now no work has addressed the problem of a running for these shapes \citep*{FE09.2,FE10.2}. As can be noted, in all non-local cases except the enfolded one we defined the level of non-Gaussianity by equating the bispectrum normalization in the equilateral limit to the same quantity for the local shape (see the discussion in \citealt{FE09.2}).

\section{Halo mass function and linear bias}\label{sct:mf}

Primordial non-Gaussianity produces modifications in the statistics of density peaks, resulting in differences in the mass function of cosmic objects and the bias of dark matter halos with respect to the underlying smooth density field. In the following we summarize how these modifications have been taken into account in the present work.

\subsection{Mass function}

For the non-Gaussian modification to the mass function of cosmic objects we adopted the prescription of \cite{LO08.1}. The main assumption behind it is that the effect of primordial non-Gaussianity on the mass function is independent of the prescription adopted to describe the mass function itself. This means that, if $n^\mathrm{(G)}_\mathrm{PS}(M,z)$ and $n_\mathrm{PS}(M,z)$ are the non-Gaussian and Gaussian mass functions, respectively, computed according to the \cite{PR74.1} recipe, we can define a correction factor $\mathcal{R}(M,z)\equiv n^\mathrm{(G)}_\mathrm{PS}(M,z)/n_\mathrm{PS}(M,z)$. Then, the non-Gaussian mass function computed according to an arbitrary prescription, $n(M,z)$ can be related to its Gaussian counterpart through

\begin{equation}
n(M,z) = \mathcal{R}(M,z) n^\mathrm{(G)}(M,z).
\end{equation}

In order to compute $n_\mathrm{PS}(M,z)$, and hence $\mathcal{R}(M,z)$, \citet{LO08.1} performed an Edgeworth expansion \citep{BL98.1} of the probability distribution for the smoothed density fluctuations field, truncating it at the linear term in $\sigma_M$. The resulting \cite{PR74.1}-like mass function reads

\begin{figure*}
	\includegraphics[width=0.45\hsize]{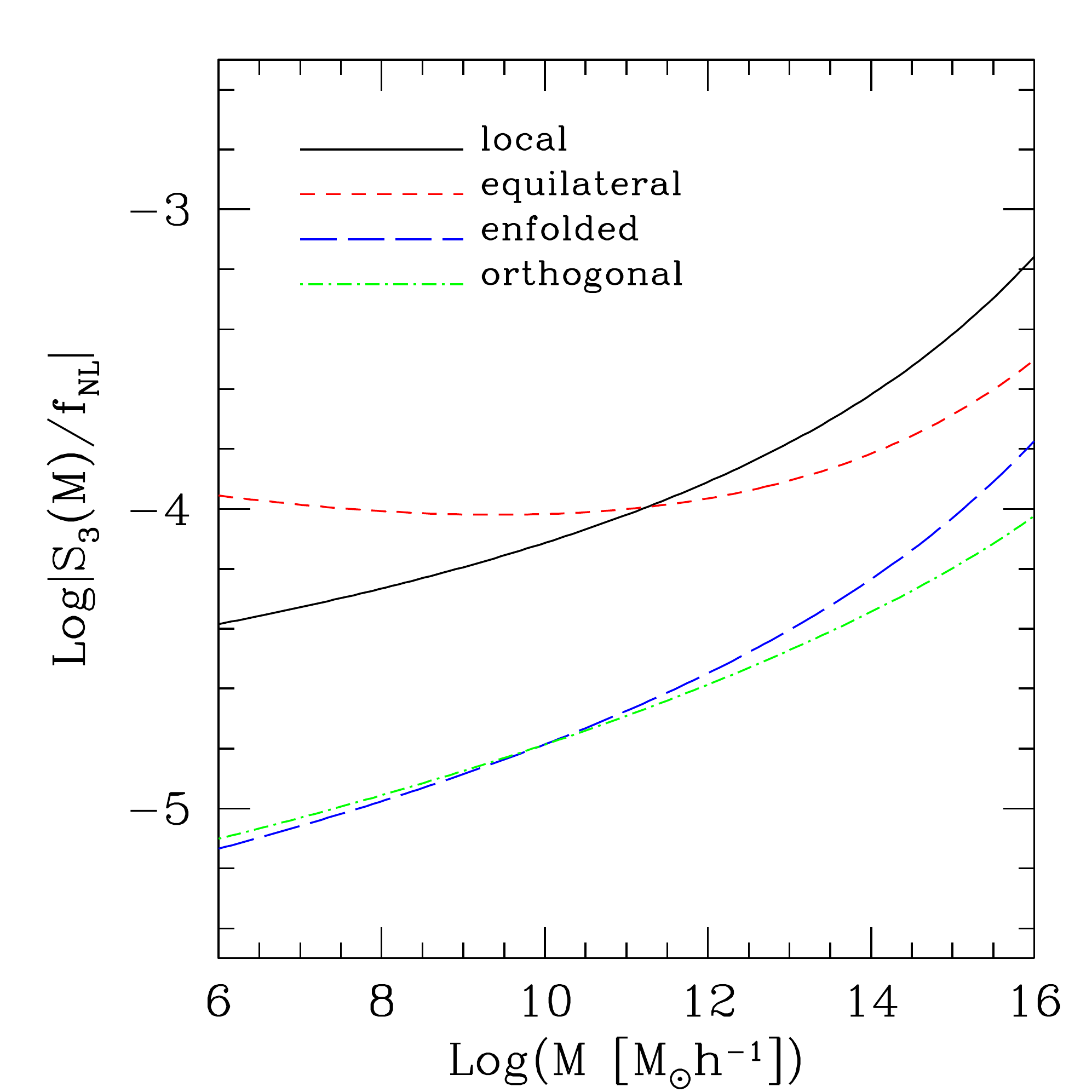}
	\includegraphics[width=0.45\hsize]{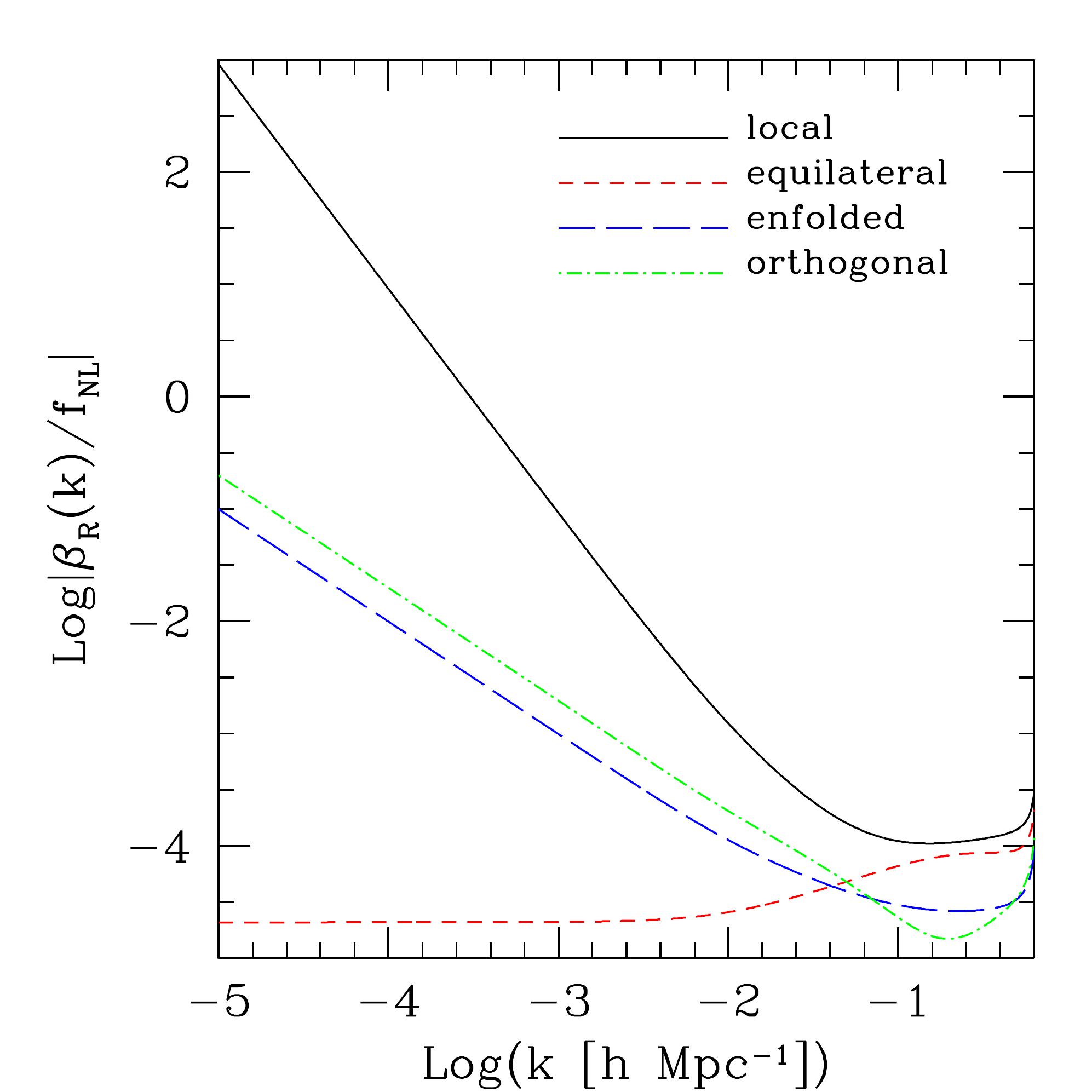}\hfill
	\caption{\emph{Left panel}. The normalized skewness in units of $f_\mathrm{NL}$ as a function of mass, for different bispectrum shapes as labeled. In the orthogonal shape case the skewness for $f_\mathrm{NL}>0$ is negative, thus the absolute value is plotted. \emph{Right panel}. The scale-dependent part of the non-Gaussian correction to the linear bias in units of $f_\mathrm{NL}$, for a $10^{14} M_\odot h^{-1}$ halo mass. Different line types refer to different shapes of the primordial bispectrum, as labeled. Note that the correction for the orthogonal shape is negative (for positive $f_\mathrm{NL}$), thus we plotted the absolute value. As specified in the text, here and for all subsequent calculations the equilateral bispectrum shape comprehend the running of $f_\mathrm{NL}$.}
\label{fig:correction}
\end{figure*}

\begin{eqnarray}\label{eqn:mfps}
n_\mathrm{PS}(M,z) &=& - \sqrt{\frac{2}{\pi}} \frac{\rho_\mathrm{m}}{M} \exp\left[ -\frac{\delta_\mathrm{c}^2(z)}{2\sigma_M^2} \right] \left[ \frac{d\ln \sigma_M}{dM} \left( \frac{\delta_\mathrm{c}(z)}{\sigma_M} + \right.\right.
\nonumber\\
&+& \left. \left. \frac{S_3\sigma_M}{6} \left( \frac{\delta_\mathrm{c}^4(z)}{\sigma^4_M} -2\frac{\delta^2_\mathrm{c}(z)}{\sigma^2_M} -1\right) \right) + \right.
\nonumber\\
&+& \left. \frac{1}{6} \frac{dS_3}{dM}\sigma_M \left( \frac{\delta^2_\mathrm{c}(z)}{\sigma^2_M} -1\right) \right].
\end{eqnarray}
In the previous equation $\rho_\mathrm{m}=3H_0^2\Omega_{\mathrm{m},0}/8\pi G$ is the comoving matter density in the Universe, $\sigma_M$ is the \emph{rms} of density fluctuations smoothed on a scale corresponding to the mass $M$, and $\delta_\mathrm{c}(z) = \Delta_\mathrm{c}/D_+(z)$, where $D_+(z)$ is the growth factor and $\Delta_\mathrm{c}$ is the critical linear overdensity for collapse. The function $S_3(M)\equiv \mu_3(M)/\sigma_M^4$ is the reduced skewness of the non-Gaussian distribution, and the skewness $\mu_3(M)$ can be computed as

\begin{eqnarray}
\mu_3(M) &=& \int_{\mathbb{R}^9} \mathcal{M}_R(k_1) \mathcal{M}_R(k_2) \mathcal{M}_R(k_3) \times
\nonumber\\
&\times&\langle\Phi({\bf k}_1)\Phi({\bf k}_2)\Phi({\bf k}_3)\rangle \frac{d{\bf k}_1d{\bf k}_2d{\bf k}_3}{(2\pi)^9}.
\end{eqnarray}

The last thing that remains to be defined is the function $\mathcal{M}_R(k)$, that relates the density fluctuations smoothed on some scale $R$ to the respective peculiar potential,

\begin{equation}
\mathcal{M}_R(k) \equiv \frac{2}{3}\frac{T(k)k^2}{H_0^2\Omega_{\mathrm{m},0}}W_R(k),
\end{equation}
where $T(k)$ is the matter transfer function and $W_R(k)$ is the Fourier transform of the top-hat window function.

In this work we adopted the \cite{BA86.1} matter transfer function, with the shape factor correction of \cite{SU95.1}. This reproduces fairly well the more sophisticated recipe of \citet{EI98.1} except for the presence of the baryon acoustic oscillation, that anyway is not of interest here. We additionally adopted as reference mass function the prescription of \citet{SH02.1} (see \citealt{JE01.1,WA06.1,TI08.1} for alternative prescriptions). Other approaches also exist for computing the non-Gaussian correction to the mass function, that give results is broad agreement with those adopted here (\citealt*{MA00.2}; \citealt{DA10.1}). These semi-analytic prescriptions are known to be in good agreement with cosmological simulations having local non-Gaussian initial conditions \citep{GR07.1,GR09.1,DE10.1}, while more recently the same has been proven to be true also for more generic primordial bispectrum shapes \citep*{WA10.1}.

In the left panel of Figure \ref{fig:correction} we show the mass dependence of the reduced skewness for the four non-Gaussian shapes that have been detailed above. It is interesting to note that in the orthogonal case the skewness for positive $f_\mathrm{NL}$ would be negative, thus bringing to a reduction in the abundance of massive objects. Also, the equilateral shape is the only case in which $S_3(M)$ is not monotonic. In computing the non-Gaussian corrections to the mass function we have taken into account the modification to the critical overdensity for collapse suggested by \citet{GR09.1} (see also \citealt*{MA10.1,MA10.3,MA10.2}), according to which $\Delta_\mathrm{c} \rightarrow \Delta_\mathrm{c}\sqrt{q}$, with $q\sim 0.8$.

\subsection{Halo bias}

Recently much attention has been devoted to the effect of primordial non-Gaussianity on halo bias, and the use thereof for constraining $f_\mathrm{NL}$ (\citealt{DA08.1,VE09.1}; \citealt*{CA08.1}). In particular, \cite{MA08.1} have shown that primordial non-Gaussianity introduce a scale dependence on the large scale halo bias. This peculiarity allows to place already stringent constraints from existing data \citep{SL08.1,AF08.1}.

The non-Gaussian halo bias can be written in a relatively straightforward way in terms of its Gaussian counterpart as \citep*{CA10.1}

\begin{equation}\label{eqn:bias}
b(M,z,k) = b^\mathrm{(G)}(M,z) + \beta_R(k)\sigma_M^2\left[ b^\mathrm{(G)}(M,z)-1 \right]^2,
\end{equation}
where the function $\beta_R(k)$ encapsulates all the scale dependence of the non-Gaussian correction to the bias, and reads

\begin{eqnarray}
\beta_R(k) &=& \frac{1}{8\pi^2\sigma_M^2\mathcal{M}_R(k)} \int_0^{+\infty} \zeta^2\mathcal{M}_R(\zeta) \times
\nonumber\\
&\times& \left[ \int_{-1}^1 \mathcal{M}_R\left(\sqrt{\alpha}\right) \frac{B_\Phi\left( \zeta,\sqrt{\alpha},k \right)}{P_\Phi(k)} d\mu \right] d\zeta,
\end{eqnarray}
where $\alpha \equiv k^2 + \zeta^2 + 2 k\zeta\mu$. In the simple case of local bispectrum shape it can be shown that the function $\beta_R(k)$ should scale as $\propto k^{-2}$ at large scales, so that a substantial boost (if $f_\mathrm{NL}>0$) in the halo bias is expected at those scales.

In the right panel of Figure \ref{fig:correction} we show the scale dependence of the function $\beta_R(k)$ for a fixed halo mass. In the remainder of this paper we adopted the \citet*{SH01.1} formula for the Gaussian bias. Thus, since we express the correction to the halo bias as a function of the Gaussian bias itself, we did not need to take into account the ellipsoidal collapse correction suggested by \citet{GR09.1}.

\section{Halo model of the large-scale structure}\label{sct:halomodel}

The halo model \citep{MA00.3,SE00.1} is a physical framework that allows the description of the correlation function of dark matter as well of different tracers of the LSS such as galaxies ad galaxy clusters. It is based on the assumption that the power spectrum of \emph{particles} (either dark matter particles or tracers) is given by the sum of two contributions: particle pairs residing in the same structure, and particle pairs residing in two different structures. The implicit hypothesis underlying this assumption is that no particles are found outside bound structures. Accordingly, the cross spectrum $P_\mathrm{xy}(k,z)$ of two different kinds of particles x and y (if x = y then $P_\mathrm{xx}(k,z)$ is the power spectrum of particle type x) can be written as the sum of two terms,

\begin{equation}
P_\mathrm{xy}(k,z) = P_{\mathrm{xy},1}(k,z)+P_{\mathrm{xy},2}(k,z).
\end{equation}
The first one is named $1-$halo term, while the second is the $2-$halo term. From the discussion above, it is immediately obvious that the first term dominates on very small scales, while the second is dominant at large scales.

The exact form of the two terms depend on the exact kind of \emph{particle} that we are considering. However, a common feature is that the $2-$halo term should depend upon the bias of dark matter halos, since it should represents particle pairs residing in separated structures \citep{CO02.2}.

\subsection{Galaxy spectrum}

In this case we set x $=$ y $=$ g, and the terms contributing to the galaxy power spectrum can be written as

\begin{equation}\label{eqn:1hg}
P_{\mathrm{gg},1}(k,z) = \int_0^{+\infty} n(M,z) \frac{\langle N_\mathrm{g}(N_\mathrm{g}-1)|M \rangle}{n_\mathrm{g}^2(z)}\left|u_\mathrm{g}(M,z,k)\right|^p dM
\end{equation}
and

\begin{equation}\label{eqn:2hg}
P_{\mathrm{gg},2}(k,z) = P_\mathrm{L}(k,z)\varphi^2_\mathrm{g}(k,z),
\end{equation}
where

\begin{equation}\label{eqn:phig}
\varphi_\mathrm{g}(k,z) = \int_0^{+\infty} n(M,z) b(M,z,k)\frac{\langle N_\mathrm{g}|M \rangle}{n_\mathrm{g}(z)} u_\mathrm{g}(M,z,k) dM.
\end{equation}
Note that we have included a scale dependence in the halo bias, in order to account for the effect of primordial non-Gaussianity. The functions $\langle N_\mathrm{g}|M \rangle$ and $\langle N_\mathrm{g}(N_\mathrm{g}-1)|M \rangle$ are the first and second moment respectively of the halo occupation distribution $p(N_\mathrm{g},M)$, which represents the probability that $N_\mathrm{g}$ galaxies reside inside a dark matter halo of mass $M$. The function $P_\mathrm{L}(k,z)=Ak^nD^2_+(z)T^2(k)$ is the linear matter power spectrum. 

We observe that the previous integrals do not actually extend down to $M=0$, since there is a minimum halo mass $M_\mathrm{g}$ below which no galaxy formation is possible, i.e., $p(N_\mathrm{g},M)=0$ for $M<M_\mathrm{g}$. Additionally, not all galaxies at all redshifts are accessible to observations, hence $M_\mathrm{g}$ should be the minimum galaxy halo mass that, at a given redshift, enters in our fiducial catalogue. We come back on the issue of the minimum mass further below. The function $n_\mathrm{g}(z)$ is the mean number density of galaxies, and reads

\begin{equation}\label{eqn:gz}
n_\mathrm{g}(z) = \int_0^{+\infty} n(M,z)\langle N_\mathrm{g}|M\rangle dM.
\end{equation}
The quantity $u_\mathrm{g}(M,z,k)$ in Eqs. (\ref{eqn:1hg}) and (\ref{eqn:phig}) represents instead the Fourier transform of the galaxy number density inside dark matter halos of mass $M$ and redshift $z$. We set it equal to the number density of dark matter particles, that is $u_\mathrm{g}(M,z,k) \equiv \hat\rho(M,z,k)/M$. For the dark matter density profile we adopt a \citet*{NA96.1} shape (NFW henceforth, see also \citealt*{NA95.1,NA97.1}) with a concentration-mass relation that reproduces simulated matter power spectra (see \citealt{SE03.2,FE10.1} for discussions). We discuss more on the adopted density profile in Section \ref{sct:discussion}.

We make the assumption that if a dark matter halo hosts at least one galaxy, then one galaxy should sit at the center of the halo itself. As a consequence, the $1-$halo term of the galaxy power spectrum consists of two contributions: galaxy pairs that involve the central objects and all the other pairs. As discussed by \citet{CO02.2} we can self-consistently assign $p=1$ to the first class ($\langle N_\mathrm{g}(N_\mathrm{g}-1)|M \rangle <1$) and $p=2$ to the second ($\langle N_\mathrm{g}(N_\mathrm{g}-1)|M \rangle \ge1$). As for the moments of the halo occupation distribution we adopted simple fitting forms that reproduce the results of semi-analytic galaxy formation models, that is

\begin{equation}\label{eqn:m1}
\langle N_\mathrm{g}|M \rangle = N_{\mathrm{g},0}\left( \frac{M}{M_0} \right)^\theta,
\end{equation}
and

\begin{equation}\label{eqn:m2}
\langle N_\mathrm{g}(N_\mathrm{g}-1)|M \rangle = f^2(M)\langle N_\mathrm{g}|M \rangle^2
\end{equation}
\citep{CO02.2,CO04.3,HU08.1}, where $f(M) = \log\sqrt{M/10^{11}M_\odot h^{-1}}$ if $M\le 10^{13} M_\odot h^{-1}$ and $f(M)=1$ otherwise. This corresponds at assuming a Poisson distribution at high halo masses with deviations thereof occurring at low masses.

\begin{figure}
	\includegraphics[width=\hsize]{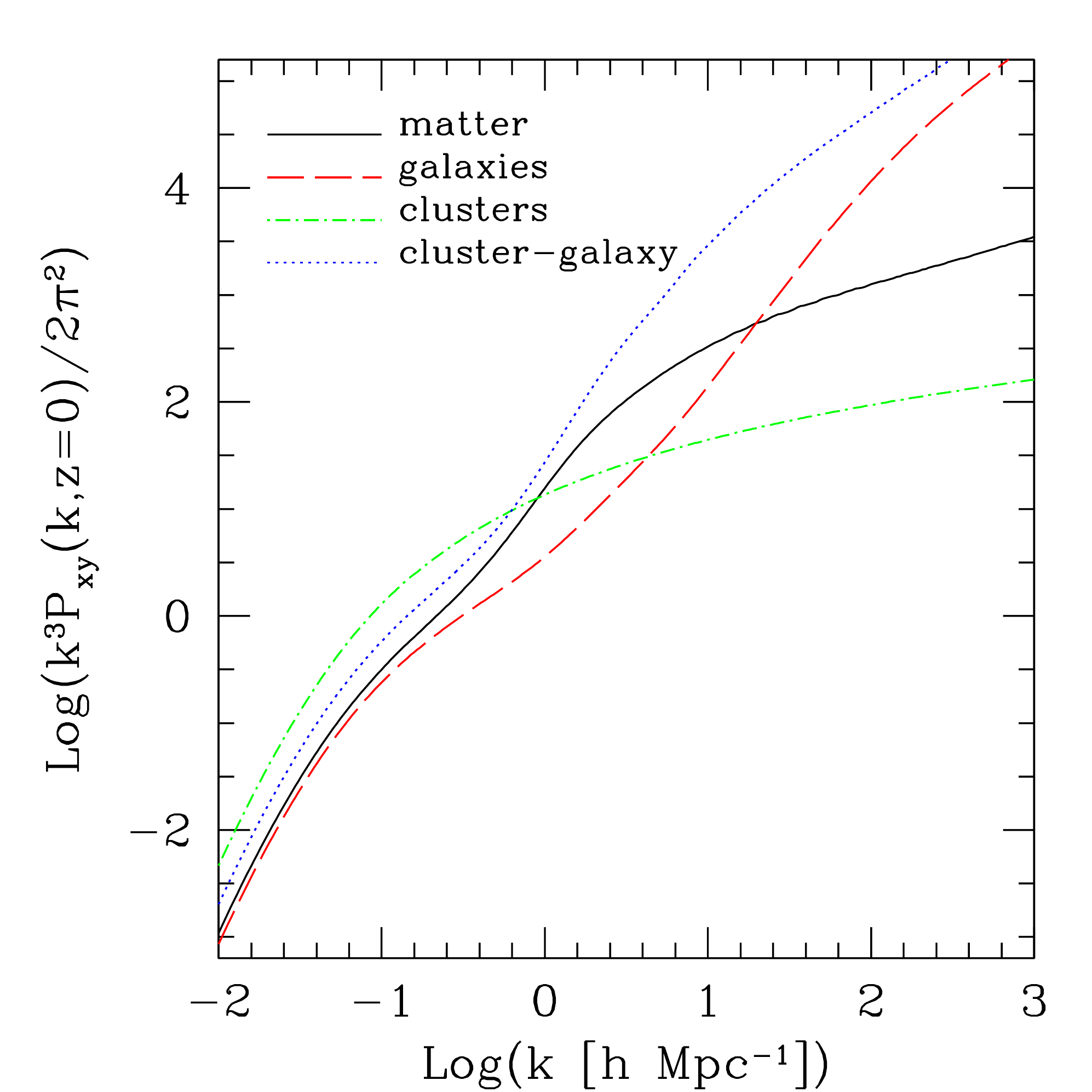}\hfill
	\caption{The power spectra of galaxies and galaxy clusters obtained according to our halo model at $z=0$, compared to the matter power spectrum and to the cluster-galaxy cross spectrum at the same redshift, as labeled.}
\label{fig:auto}
\end{figure}

The three free parameters $N_{\mathrm{g},0}$, $M_0$ and $\theta$ depend on the type of galaxy that is under consideration \citep{CO02.2}. Typically, for blue galaxies $N_{\mathrm{g},0} = 0.7$, $M_0 = 4\times 10^{12} M_\odot h^{-1}$, $\theta=0$ for $M\le M_0$ and $\theta=0.8$ otherwise. For red galaxies we have instead $N_{\mathrm{g},0}=1$, $M_0=2.5\times 10^{12}M_\odot h^{-1}$, and $\theta=0.9$. Obviously, the average number of galaxies inside a halo of some mass $M$ is the sum of the average number of red galaxies and of blue galaxies. Now, the distinction between blue and red galaxies can be rather qualitative, and dependent on the framework we are interested in (e.g. observations vs. semi-analytic models). The slitless spectroscopic instrument planned for \emph{Euclid} will be mainly sensitive to the flux and equivalent width of emission lines, mainly H$\alpha$, and thus will mostly observe star-forming galaxies. Although this category includes dust-obscured starburst objects, that are not properly blue, we believe the parameter set for blue galaxies to be more suited for the target of \emph{Euclid}. Hence we shall adopt it in the remainder of this work. In Section \ref{sct:discussion} we discuss how our results are changed if we consider a multi-slit spectroscopic instrument, that select a different galaxy population.

\subsection{Cluster spectrum}

For galaxy clusters the situation is less elaborated, since it is commonly assumed that only one cluster is contained inside each dark matter halo that is massive enough (this is obviously not true for galaxies and dark matter particles). Consequently, the $1-$halo term vanishes, $P_{\mathrm{cc},1}(k,z)=0$, while the $2-$halo term reads

\begin{equation}
P_{\mathrm{cc},2}(k,z) = P_\mathrm{L}(k,z) \varphi_\mathrm{c}^2(k,z),
\end{equation}
where

\begin{equation}
\varphi_\mathrm{c}(k,z) = \int_0^{+\infty} n(M,z) \frac{b(M,z,k)}{n_\mathrm{c}(z)}dM.
\end{equation}

In this case as well there is a minimum halo mass $M_\mathrm{c}$ below which no cluster is formed (or, below which clusters are not observable), and the integrals are to be evaluated above that mass. The function $n_\mathrm{c}(z)$ then reads

\begin{equation}\label{eqn:cz}
n_\mathrm{c}(z) = \int_{0}^{+\infty} n(M,z) dM.
\end{equation}
Further below we detail how the issue of the minimum cluster mass is addressed in this paper.

In Figure \ref{fig:auto} we show the power spectrum of galaxy clusters alongside that for galaxies, and compare them to the dark matter power spectrum obtained with the halo model \citep{FE10.1}. At linear scales clusters are more biased than galaxies as is to be expected. Moreover, since the power spectrum of clusters is made only by the $2-$halo term, their correlation drops off substantially at small scales as compared to other tracers.

\subsection{Cluster-galaxy cross spectrum}

Extending the previous results, \citet{HU08.1} derived the two halo model contributions to the cross correlation of clusters and galaxies. They can be written, respectively, as

\begin{equation}
P_{\mathrm{cg},1}(k,z) = \int_{0}^{+\infty} n(M,z) \frac{\langle N_\mathrm{g}|M \rangle}{n_\mathrm{g}(z)n_\mathrm{c}(z)}u_\mathrm{g}(M,z,k) dM
\end{equation}
and

\begin{equation}
P_{\mathrm{cg},2}(k,z) = P_\mathrm{L}(k,z)\varphi_\mathrm{c}(k,z)\varphi_\mathrm{g}(k,z).
\end{equation}

\begin{figure}
	\includegraphics[width=\hsize]{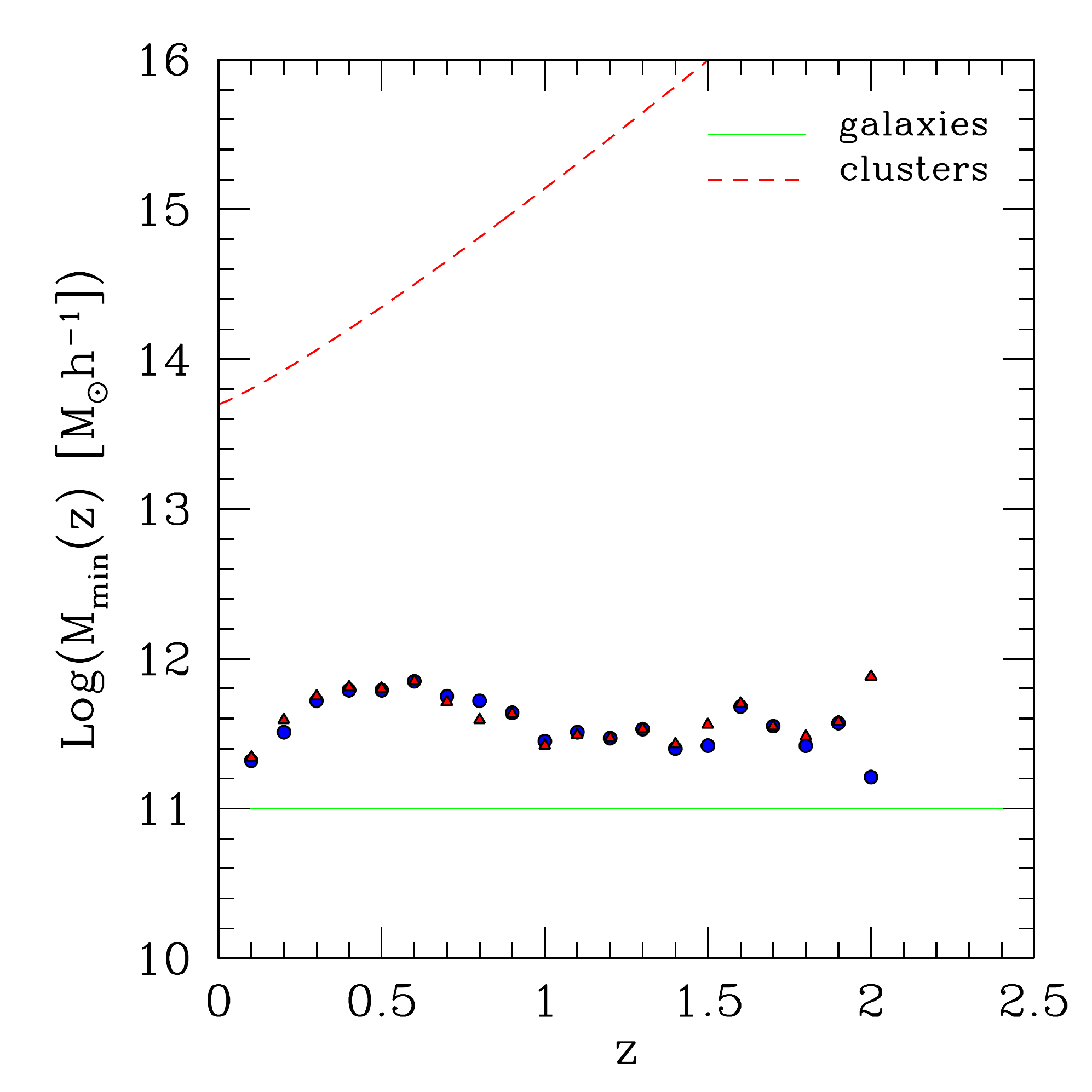}\hfill
	\caption{The minimum mass as a function of redshift for spectroscopically selected galaxies (solid green line) and weak lensing selected clusters (dashed red line) adopted in the present work. The points are the average galaxy halo masses derived from semi-analytic evolutionary models for two different H$\alpha$ flux thresholds, $\log(f_{H\alpha}) = -15.4$ (red triangles) and $\log(f_{H\alpha}) = -15.5$ (blue circles, CGS units).}
\label{fig:mMin}
\end{figure}

In Figure \ref{fig:auto} we also show the cross spectrum of galaxy clusters and galaxies. As can be seen, at large scales the cross spectrum stays in between the spectra of galaxies and clusters. However, it shows more power on small scales than both clusters and galaxies. The fact that the scale dependence of the cross spectrum is different from that of the galaxy spectrum and cluster spectrum means that it carries a different kind of information on the spatial distribution of LSS tracers, thus the cross-correlation can make up an important improvement when constraining cosmological parameters.

\subsection{Minimum masses}

As mentioned above, in order to compute the cluster and galaxy power spectra it is necessary to select a minimum mass for both classes of objects. In general this minimum mass depends on redshift. Let us start with clusters. \citet*{BE10.2} have estimated the weak lensing selection function for galaxy clusters given a \emph{Euclid}-like survey and different S/N detection thresholds (see also \citealt{LA09.1}). In order to be conservative, we adopted S/N $\ge5$, and fitted the corresponding curve in Figure 1 of \citet*{BE10.2}. The result is displayed by the red dashed curve in Figure \ref{fig:mMin}, which represents the $M_\mathrm{c}(z)$ we adopted in this work.

\begin{figure}
	\includegraphics[width=\hsize]{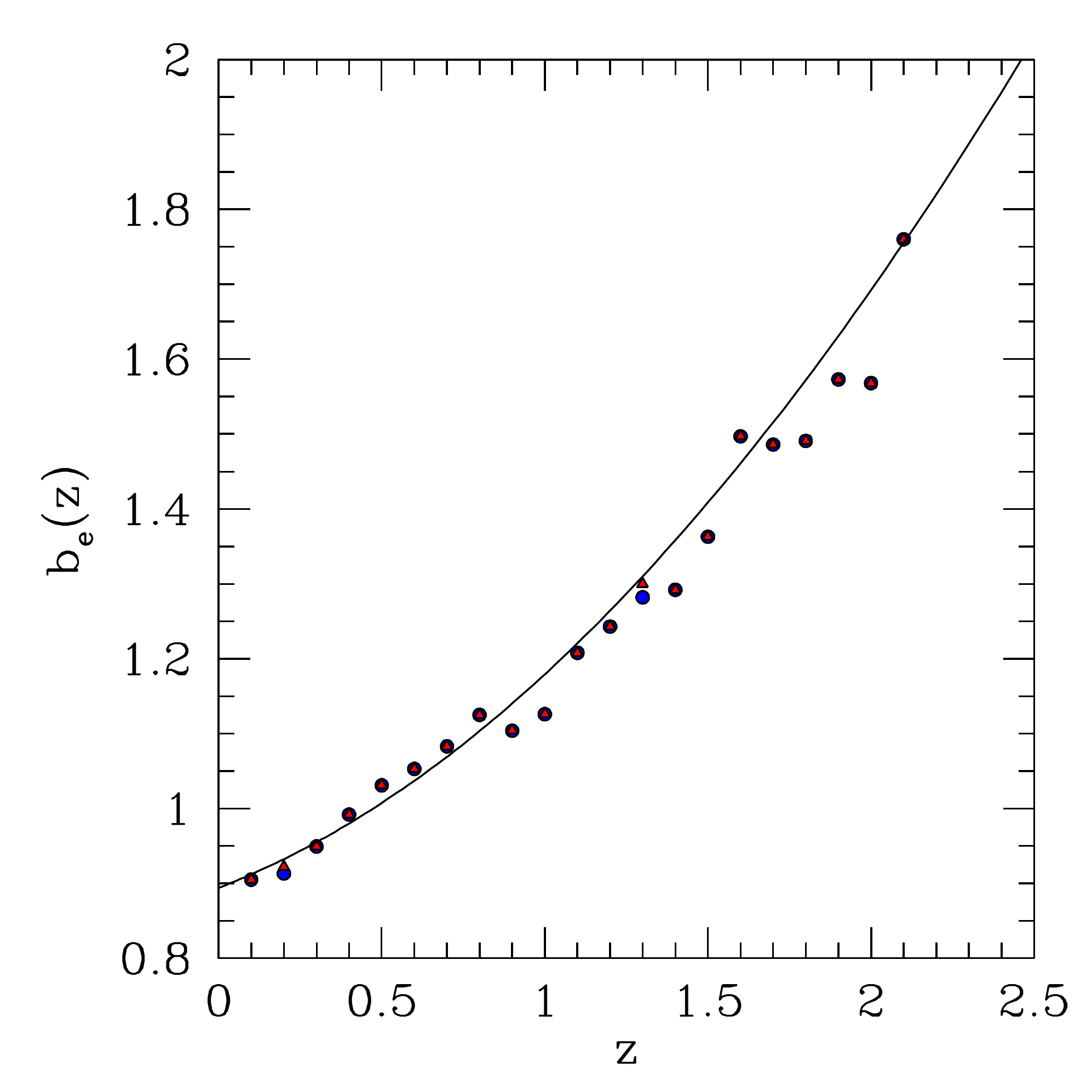}\hfill
	\caption{The effective bias as a function of redshift for spectroscopically selected galaxies (solid black line). The points are the galaxy biases derived from semi-analytic evolutionary models for two different H$\alpha$ flux thresholds, $\log(f_{\mathrm{H}\alpha}) = -15.4$ (red triangles) and $\log(f_{\mathrm{H}\alpha}) = -15.5$ (blue circles, CGS units).}
\label{fig:bGal}
\end{figure}

For galaxies, we adopted the minimum halo mass $M_\mathrm{g}(z) = 10^{11} M_\odot h^{-1}$ which, given the halo occupation moments reported in Eqs. (\ref{eqn:m1}) and (\ref{eqn:m2}), reproduces in the reference Gaussian cosmology the effective bias predicted for H$\alpha$ selected galaxies by \emph{Euclid} \citep{OR10.1}. According to Eq. (\ref{eqn:2hg}), the effective galaxy bias can be estimated within the framework of the halo model as the large scale limit of $\varphi_\mathrm{g}(k,z)$, i.e. $b_\mathrm{e}(z) \equiv \lim_{k \to 0} \varphi_\mathrm{g}(k,z)$. The effective bias estimates reported in \citet{OR10.1} have been performed adopting semi-analitic models of galaxy formation \citep{BA05.1,BO06.1}, and assuming a spectroscopic selection based on H$\alpha$ emission. The latter is appropriate for our scope, since we are considering star-forming galaxies only. The points reported in Figure \ref{fig:bGal} are the predictions performed by \citet{OR10.1} using the \citet{BA05.1} model for two different H$\alpha$ flux thresholds ($\log(f_{\mathrm{H}\alpha}) = -15.4$ and $\log(f_{\mathrm{H}\alpha}) = -15.5$ in CGS units). Obviously there is not much difference between the two choices, that are well reproduced by the solid line, representing the outcome of our halo model. In Figure \ref{fig:mMin} we also report the minimum mass of galaxy halos, compared to the average such mass found again by \citet{OR10.1}.
 
\section{Results}\label{sct:results}

In this Section we summarize our main results. First, we explored what happens to the power and cross spectra described above when the distribution of primordial density fluctuations deviates from a Gaussian. Then, we performed a Fisher matrix analysis in order to forecast the constraints on the main cosmological parameters that can be expected by a \emph{Euclid}-like all-sky survey.

\subsection{Effect of non-Gaussianity on the power and cross spectra}

\begin{figure*}
	\includegraphics[width=0.85\hsize]{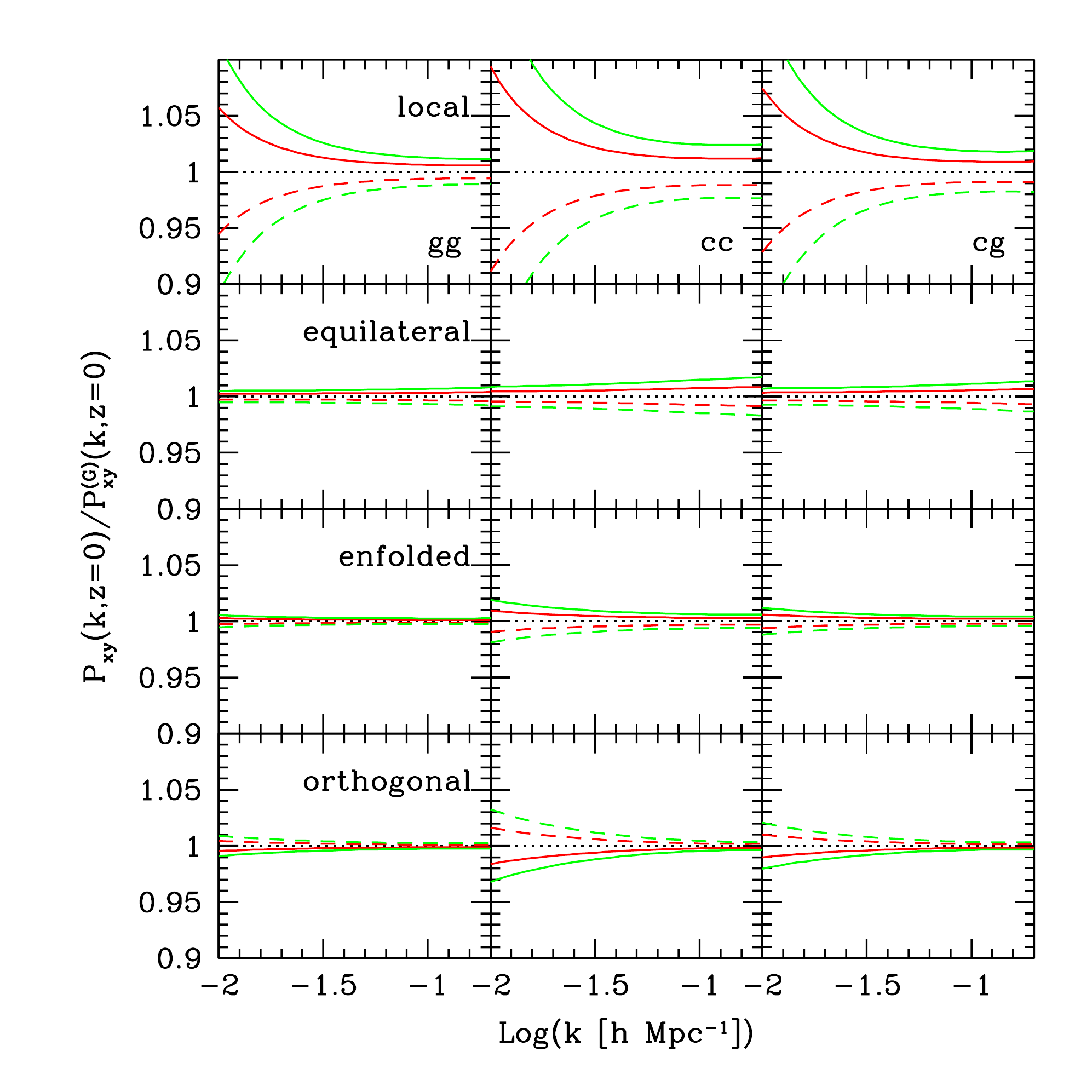}\hfill
	\caption{The ratio of the power and cross spectra for different tracers of the LSS computed in various non-Gaussian cosmologies to the same quantities evaluated in the reference Gaussian case. From top to bottom: non-Gaussianities with local, equilateral, enfolded, and orthogonal bispectrum shapes. From left to right: the power spectrum of star-forming galaxies, galaxy clusters, and the cross spectrum of clusters and galaxies. For all non-Gaussian cosmologies considered here values of $f_\mathrm{NL}=100$ (solid red lines), $f_\mathrm{NL}=200$ (solid green lines), $f_\mathrm{NL}=-100$ (dashed red lines), and $f_\mathrm{NL}=-200$ (dashed green lines) were chosen. All results are shown at $z=0$.}
\label{fig:ngSpectrum}
\end{figure*}

As a first step of our analysis we computed the \emph{Euclid}-expected galaxy power spectrum, cluster power spectrum, and cluster-galaxy cross spectrum for all the cosmological models considered in this work, and investigated the modifications induced by primordial non-Gaussianity on each one. In Figure \ref{fig:ngSpectrum} we show the effect of the different shapes of primordial non-Gaussianity examined in the present paper on the power and cross spectra for different tracers of the LSS. We show results for $f_\mathrm{NL}=\pm 100$, since this roughly corresponds to the best high confidence limits that are currently placed on this parameter, and for $f_\mathrm{NL}=\pm 200$ as a comparison. We recall that the power spectrum of dark matter halos in cosmological simulations with local non-Gaussian initial conditions has been studied in \citet{GR09.1} and \citet*{DE09.1}. However, here we are interested in a wider variety of bispectrum shapes, and in specific classes of objects that are not necessarily in a one-to-one relation with dark matter halos.

The corrections to the power and cross spectra that are due to different kinds of primordial non-Gaussianity resemble the corrections to the large scale halo bias that have been shown in Figure \ref{fig:correction}. This is to be expected, since in the halo model the mass function is normalized over (see Section \ref{sct:halomodel}), hence the largest effect is expected to be due to the halo bias. In particular, for the local shape we observe a sharp increase of the power at large scales for all tracers. A similar increase in power at large scales, although not as strong, is also observed for the enfolded and orthogonal shapes. Note also that the trend for the latter is reversed, due to the fact that in this case a positive $f_\mathrm{NL}$ implies a negative effect on the bias and the mass function. Finally, for the equilateral shape we find a slight increase in the power at intermediate scales, again in agreement with the behavior of the halo bias.

As one might expect, the larger effect is found in correspondence of the tracers that are more biased with respect to the underlying matter density field, that are galaxy clusters. This can be understood by looking at Eq. (\ref{eqn:bias}), where it is shown that the difference between the non-Gaussian halo bias and its Gaussian counterpart is in fact proportional to the square of the Gaussian bias itself. On the other hand, the effect on the cluster-galaxy cross power spectrum is at an intermediate level between the effect on the galaxy power spectrum and that on the cluster power spectrum. Due to their low average bias, star-forming galaxies are very poorly affected by primordial non-Gaussianity, and this is particularly true for the enfolded and orthogonal bispectrum shapes.

A closer look to the first and third column panels of Figure \ref{fig:ngSpectrum} reveals an interesting fact. While the change in the cluster-galaxy cross spectrum due to primordial non-Gaussianity is larger than the change in the galaxy power spectrum, it is more so for the enfolded and orthogonal bispectrum shapes, as compared to the local and equilateral ones. This means that the inclusion of the cross-correlation between clusters and galaxies should bring a better improvement over the constraints on non-Gaussianity obtained by using the galaxy spectrum alone in the first two cases than in the second ones. Although it is difficult to gauge what the effect of the additional inclusion of the cluster power spectrum would be, we show in the next subsection that this expectation is to some extent satisfied.

\subsection{Fisher matrix analysis}

For the Fisher information matrix of the cross correlation between clusters and galaxies we followed the calculations of \citet{HU08.1}, assuming perfect redshift knowledge for both galaxies and clusters. This assumption is justified since galaxies are spectroscopically selected, while clusters, being chosen as the highest S/N cosmic shear peaks, should be very massive, thus making follow-up confirmations relatively straightforward. In addition, we assumed redshift space distortions to be sufficiently well understood to be modeled away. Accordingly, the Fisher matrix for the redshift bin centered at $z$ can be well approximated as

\begin{equation}\label{eqn:fisher}
\mathcal{F}_\mathrm{cg}^{ij}(z) \simeq \frac{V(z)}{4\pi^2}\int_{k_\mathrm{min}}^{k_\mathrm{max}} k^2dk\,w_\mathrm{cg}(k,z)
\frac{\partial \ln P_\mathrm{cg}(k,z)}{\partial \xi_i} \frac{\partial \ln P_\mathrm{cg}(k,z)}{\partial \xi_j},
\end{equation} 
where $w_\mathrm{cg}(k,z) = 2/\left[1+\gamma_\mathrm{cg}(k,z)\right]$, and

\begin{table*}
\caption{The adopted parameter set for our fiducial \emph{Euclid}-like survey}
\label{tab:survey}
\begin{tabular}{lcc}
\hline
\hline
&GALAXIES &GALAXY CLUSTERS\\
\hline
\hline
Sky Coverage&$20,000$ sq. deg. &$20,000$ sq. deg.\\
Redshift Efficiency&$0.5$&$-$\\
H$\alpha$ Flux Threshold&$3\times 10^{-16}$ erg s$^{-1}$ cm$^{-2}$&$-$\\
Cosmic Shear S/N Threshold& $-$ & $5$\\
Redshift Range &$[0.5,2.1]$&$[0.5,2.1]$\\
Minimum Scale &$k_\mathrm{max}=0.3$ Mpc$^{-1}$&$k_\mathrm{max}=0.3$ Mpc$^{-1}$\\
\hline
\hline
\end{tabular}
\end{table*}

\begin{equation}
\gamma_\mathrm{cg}(k,z) = \frac{[1+n_\mathrm{c}(z)P_\mathrm{cc}(k,z)][1+n_\mathrm{g}(z)P_\mathrm{gg}(k,z)]}{n_\mathrm{c}(z)n_\mathrm{g}(z)P^2_\mathrm{cg}(k,z)}.
\end{equation}

\begin{figure*}
\begin{center}
	\includegraphics[width=0.45\textwidth]{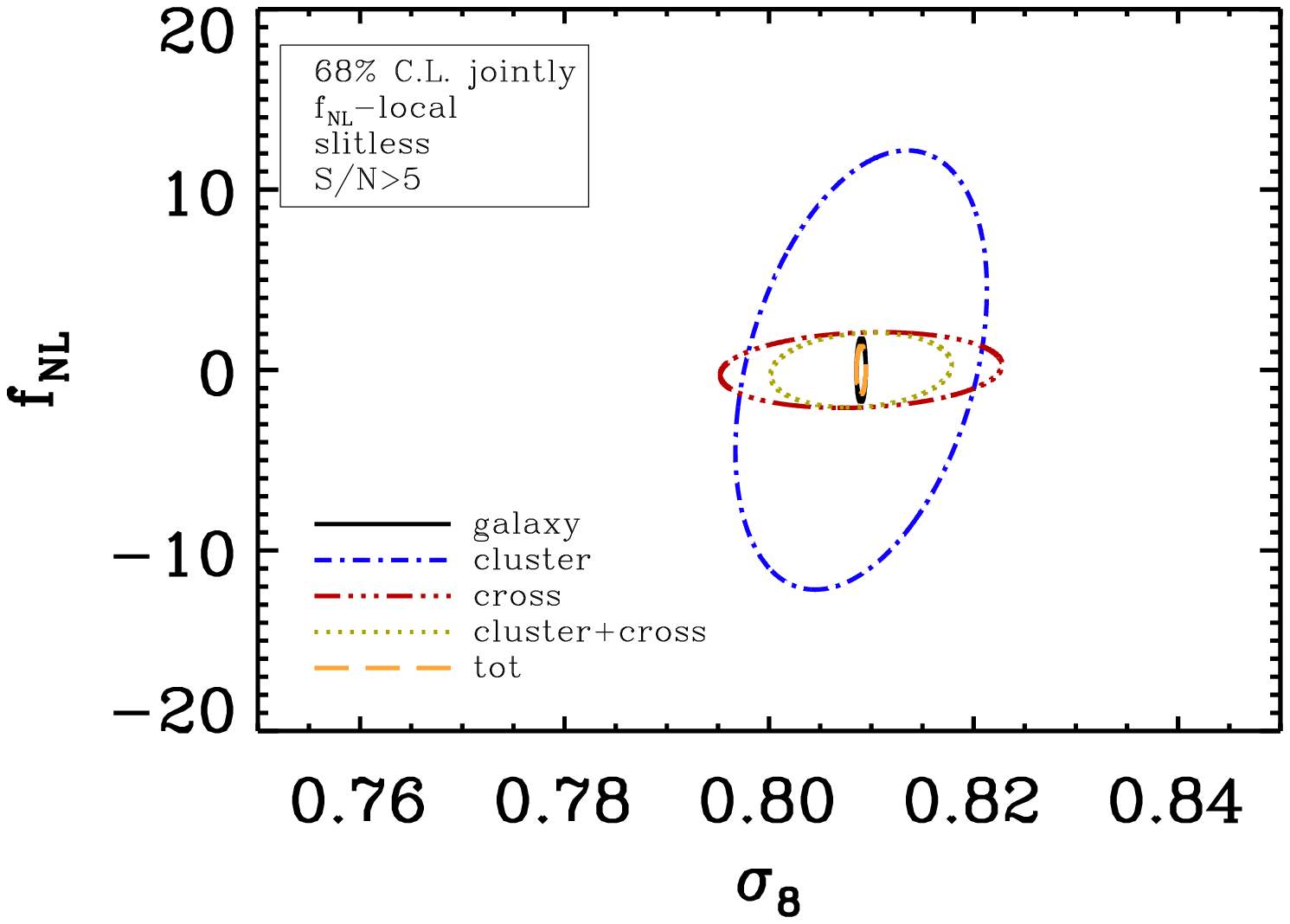}
	\includegraphics[width=0.45\textwidth]{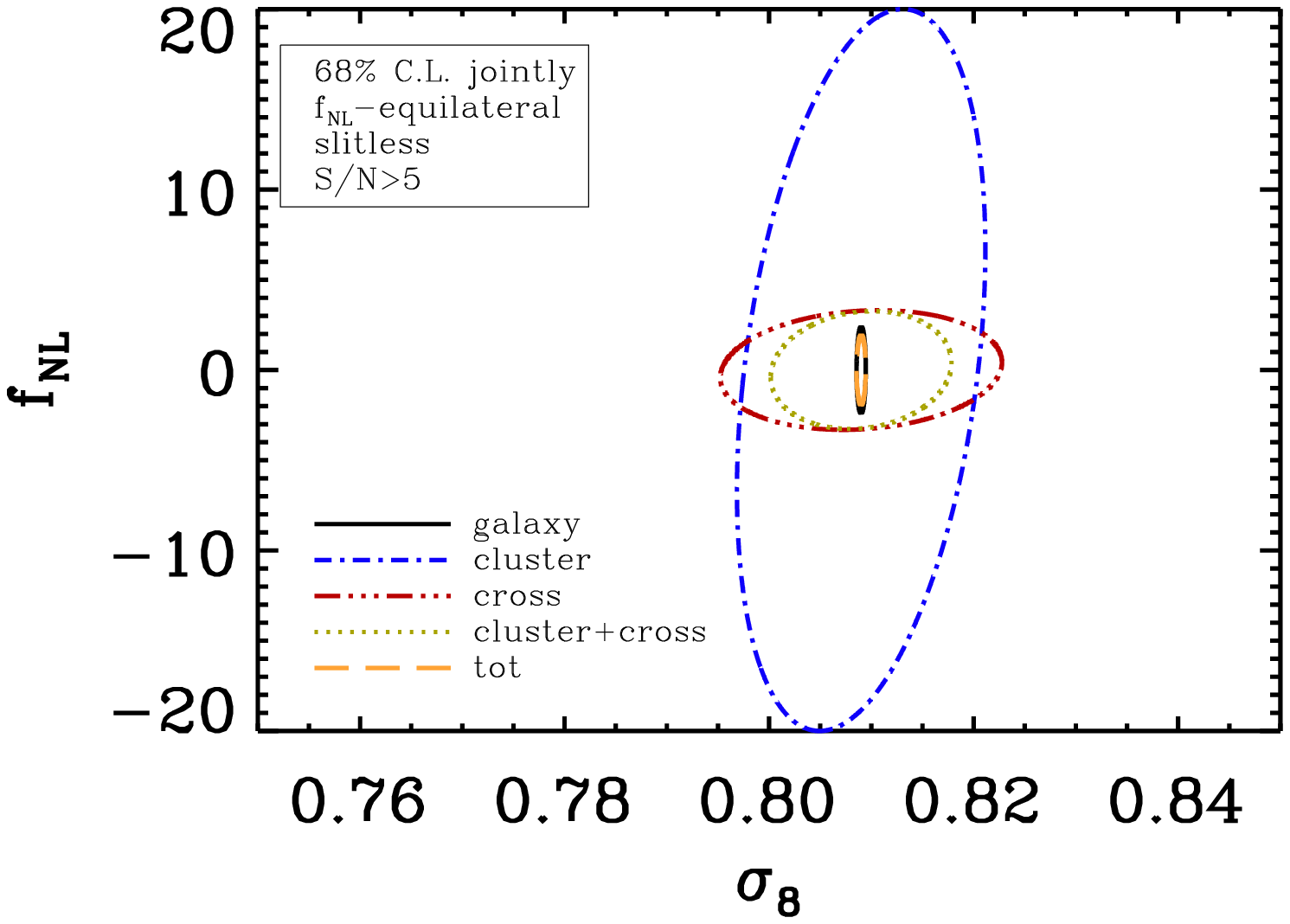}
	\includegraphics[width=0.45\textwidth]{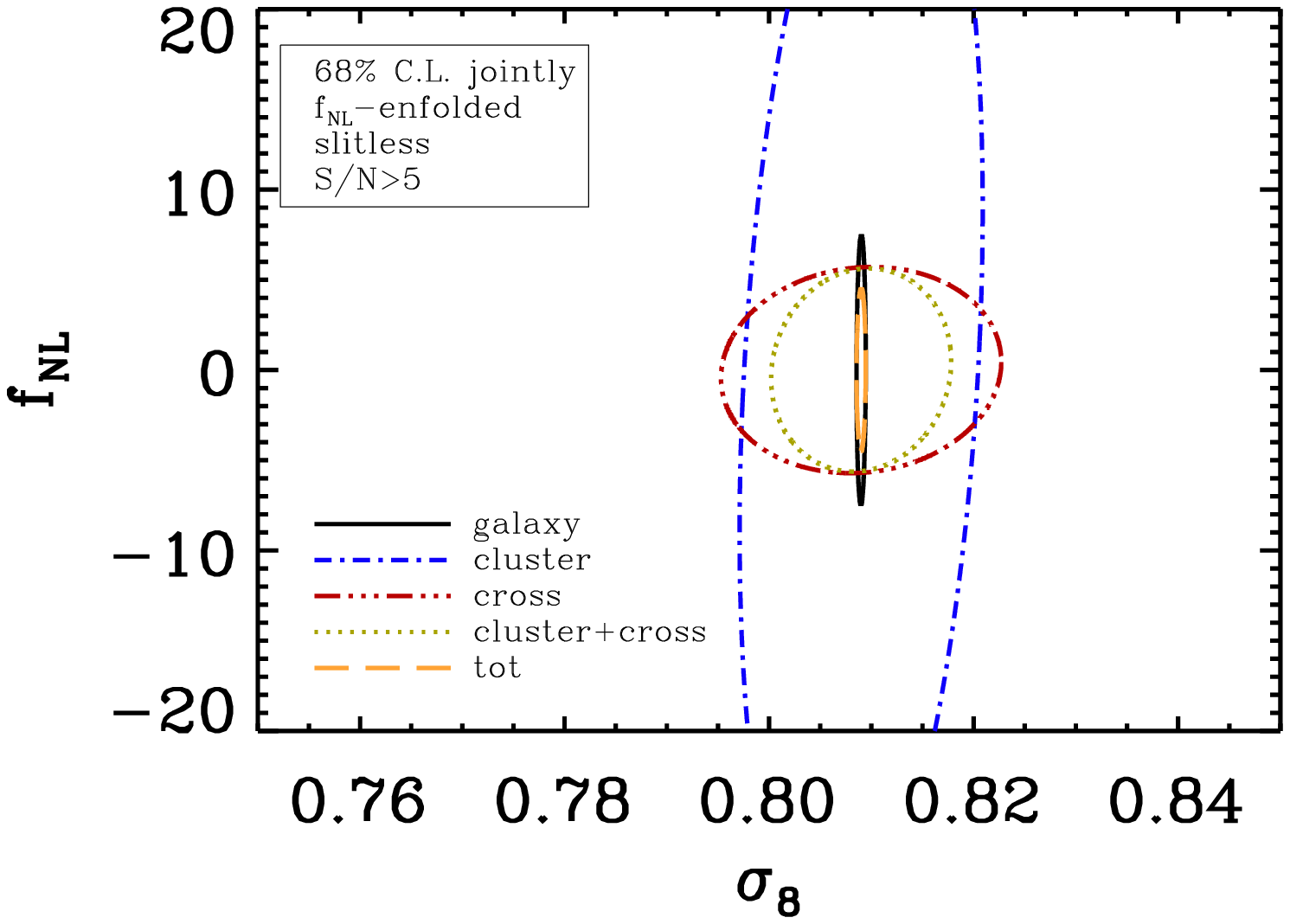}
	\includegraphics[width=0.45\textwidth]{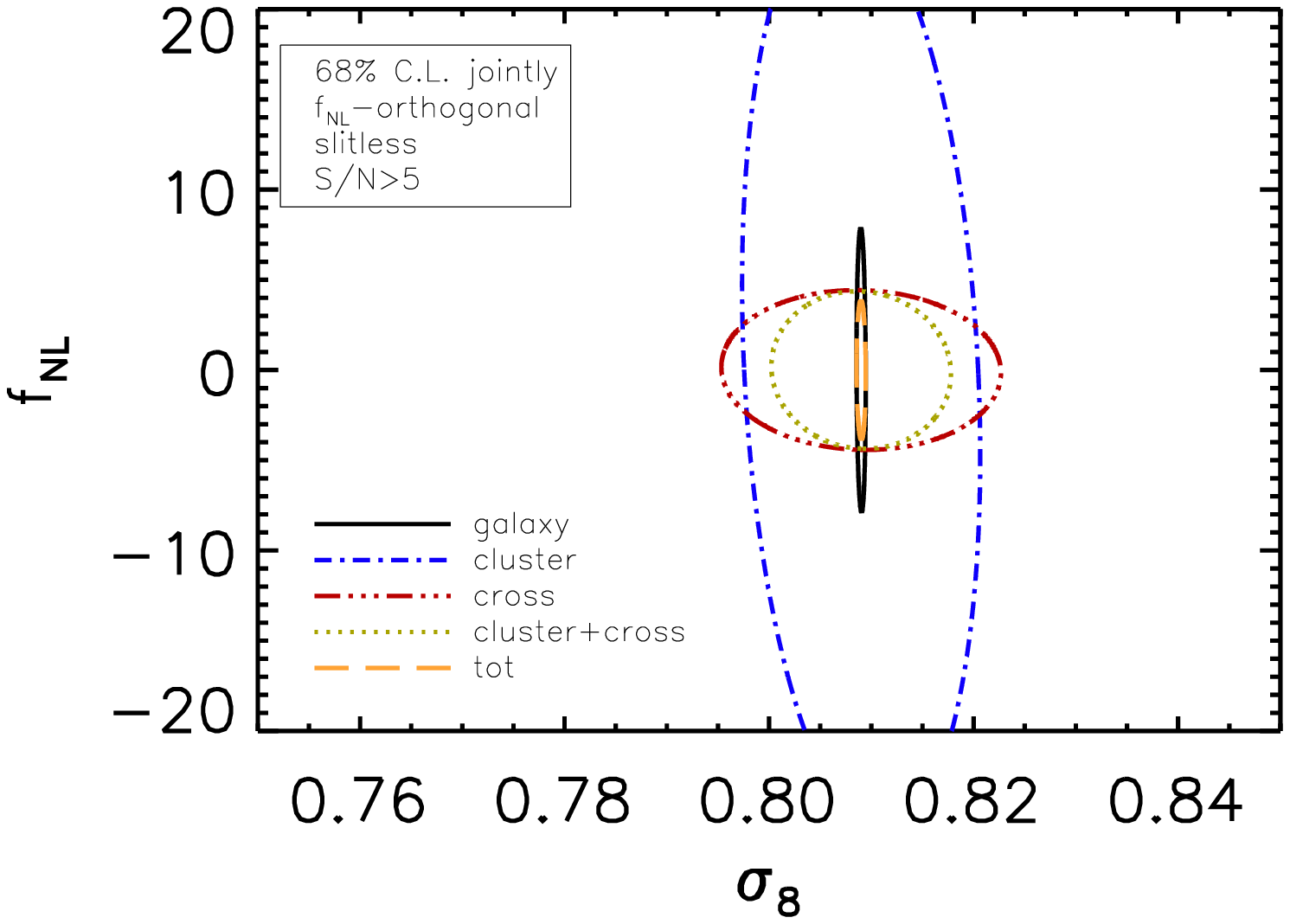}
\end{center}
\caption{The joint $68\%$ confidence level contours in the $f_\mathrm{NL}-\sigma_8$ plane given by 	different combinations of the galaxy and cluster power spectra and the cluster-galaxy cross 	spectrum, as labeled. Results are shown for all the four shapes of the primordial bispectrum that have been considered in this work. All other cosmological parameters have been kept fixed to their WMAP-7 fiducial values.}
\label{fig:ellipses}
\end{figure*}

In the above set of equations, the quantities $n_\mathrm{c}(z)$ and $n_\mathrm{g}(z)$ are the average number densities of clusters and galaxies in the survey at hand (Eqs. \ref{eqn:cz} and \ref{eqn:gz}, respectively), $V(z)$ is the comoving volume contained in the unit redshift around $z$, and $\xi_i$ is the $i-$th parameter of our cosmological model. The two wavenumbers $k_\mathrm{min}$ and $k_\mathrm{max}$ represent the boundaries of the wavenumber range used in the analysis. The chosen redshift bins should be relatively narrow, in order to treat the spatial number densities of tracers within each bin as constants.The total Fisher matrix is then given by the sum over the $n_z$ adopted redshift bins,

\begin{equation}
\mathcal{F}_\mathrm{cg}^{ij} = \sum_{\ell=1}^{n_z} \mathcal{F}_\mathrm{cg}^{ij}(z_\ell).
\end{equation}

In the case in which we are interested only in galaxies or only in clusters, the formalism remains the same, with the only difference being given by the replacements of $P_\mathrm{cg}(k,z)$ with $P_\mathrm{xx}(k,z)$ (with x $=$ c or x $=$ g) and of $w_\mathrm{cg}(k,z)$ with

\begin{equation}
w_\mathrm{xx}(k,z) = \left[ \frac{n_\mathrm{x}(z)P_\mathrm{x}(k,z)}{1+n_\mathrm{x}(z)P_\mathrm{x}(k,z)} \right]^2
\end{equation}
in Eq. (\ref{eqn:fisher}). In our analysis we adopted a redshift range typical for H$\alpha$ galaxies selected by \emph{Euclid}, $0.5 \le z \le 2.1$ (cf. the \emph{Euclid} Yellow Book), and a range of scales corresponding to $k_\mathrm{min}\le k \le 0.3\;\mathrm{Mpc}^{-1}$, where $k_\mathrm{min}$ matches the largest scale available for a given redshift bin. For completeness, in Table \ref{tab:survey} we summarize all the survey parameters that have been adopted in our Fisher matrix analysis including the redshift efficiency, that is the fraction of spectroscopically selected galaxies for which a reliable redshift measurement is expected.

\begin{table*}
\caption{The $1-\sigma$ forecasted errors for $f_\mathrm{NL}$ and $\sigma_8$.}
\label{tab:errors}
\begin{tabular}{lcccccccc}
\hline
\hline
{}&\multicolumn{2}{c}{LOCAL}&\multicolumn{2}{c}{EQUILATERAL}&\multicolumn{2}{c}{ENFOLDED}&\multicolumn{2}{c}{ORTHOGONAL}\\
\hline
\hline
{} &$\Delta f_{\rm NL}$&$\Delta\sigma_8$& $\Delta f_{\rm NL}$&$\Delta\sigma_8$& $\Delta f_{\rm NL}$&$\Delta\sigma_8$& $\Delta f_{\rm NL}$&$\Delta\sigma_8$\\
\hline
$P_\mathrm{gg}$              & $1.15$&$2.96\times 10^{-4}$ & $1.55$&$2.95\times 10^{-4}$ & $4.89$&$2.96\times 10^{-4}$ & $5.15$&$2.97\times 10^{-4}$\\
$P_\mathrm{cc}$              & $8.02$&$8.11\times 10^{-3}$ & $13.2$&$8.00\times 10^{-3}$ & $21.9$&$7.84\times 10^{-3}$ & $16.9$&$7.68\times 10^{-3}$\\
$P_\mathrm{cg}$              & $1.38$&$9.09\times 10^{-3}$ & $2.18$&$9.08\times 10^{-3}$ & $3.77$&$9.04\times 10^{-3}$ & $2.92$&$9.01\times 10^{-3}$\\
$P_\mathrm{cc}+P_\mathrm{cg}$       & $1.36$&$5.84\times 10^{-3}$ & $2.15$&$5.83\times 10^{-3}$ & $3.71$&$5.80\times 10^{-3}$ & $2.88$&$5.79\times 10^{-3}$\\
$P_\mathrm{gg}+P_\mathrm{cc}+P_\mathrm{cg}$& $0.87$&$2.95\times 10^{-4}$ & $1.25$&$2.95\times 10^{-4}$ & $2.95$&$2.95\times 10^{-4}$ & $2.51$&$2.95\times 10^{-3}$\\
\hline
\hline
\end{tabular}
\end{table*}

In Figure \ref{fig:ellipses} we show the joint $68\%$ confidence level contours in the $f_\mathrm{NL}-\sigma_8$ plane that result from the Fisher matrix analysis performed for all the four bispectrum shapes considered in the present paper. All other cosmological parameters are have been kept fixed to the fiducial WMAP-7 values summarized in Section \ref{sct:introduction}. We report the results of considering the galaxy power spectrum alone, the cluster power spectrum, the cluster-galaxy cross spectrum, and different combinations of the three. For further clarity, in Table \ref{tab:errors} we report the $1-\sigma$ errors that are forecasted both on the level of non-Gaussianity $f_\mathrm{NL}$ and on the amplitude of the matter power spectrum $\sigma_8$. The fiducial model that we adopted has $f_\mathrm{NL} = 0$ and $\sigma_8 = 0.809$, as specified in Section \ref{sct:introduction}.

As can be seen, the tracers of the LSS that individually give the best constraints on both parameters considered are galaxies, due to their very large number density which beats down the shot noise in the Fisher matrix. Galaxies alone give constraints on $\sigma_8$ that are more than an order of magnitude better than those of clusters alone, while the performance on the level of primordial non-Gaussianity $f_\mathrm{NL}$, as expected, depends on the specific choice of the primordial bispectrum shape. Galaxies perform a factor of $\sim 8$ better than clusters in the equilateral case, but only a factor of $\sim 3$ better in the orthogonal configuration. 

The constraints on the $f_\mathrm{NL}-\sigma_8$ plane that we gather from clusters alone are somewhat different than those reported in \citet{SA10.1}, which performed a similar Fisher matrix analysis on the cluster sample expected to be detected by the planned Wide Field X-ray Telescope (WFXT) (see also \citealt*{CU10.1} for a related work). Particularly, while the forecasted error on $f_\mathrm{NL}$ is comparable, the error on $\sigma_8$ is much smaller in our case than in their. We ascribe this difference to two factors. First, clusters are selected in different ways in the two cases. Second, \citet{SA10.1} introduced nuisance parameters for estimating the uncertainty in the mass calibration, and at the same time adopted priors on several cosmological parameters inspired by \emph{Planck} forecasts. We infer the second factor to be the dominant one, as rerunning the analysis of \citet{SA10.1} without nuisance parameters and priors brings to errors on $\sigma_8$ comparable with ours own (B. Sartoris, private communication). We believe it to be safe to assume that \emph{Euclid} weak lensing cluster catalogs will have sufficient control over systematics to ignore the contribution of imperfect knowledge of the mass-observable relation. This point of view is shared by \citet*{BE10.2}.

The behavior of the confidence contours given by the cluster-galaxy cross spectrum alone is interesting. Firstly, while the constraining power on $f_\mathrm{NL}$ is greatly enhanced as compared to the constraining power of clusters alone, the same is not true for $\sigma_8$. Actually the $1-\sigma$ errors on $\sigma_8$ expected from the cluster-galaxy cross correlation are systematically $\sim 15\%$ larger than those expected from clusters alone. As a consequence, while galaxies still play a dominant role about constraining $\sigma_8$, the cross correlation gives comparable constraints on $f_\mathrm{NL}$. As a matter of fact, for the enfolded and the orthogonal shapes, the cross spectrum performs even better with respect to $f_\mathrm{NL}$, so that combination of all the three probes brings to a significant improvement over galaxies alone. Specifically, while the combined probes decrease the $1-\sigma$ error on $f_\mathrm{NL}$ only by $\sim 30\%$ for the local bispectrum shape, this decrement reaches a factor of $\sim 2$ in the orthogonal case. The fact that the improvement in constraining power due to the inclusion of the cross spectrum is more pronounced for the enfolded and orthogonal shapes can be deduced from the observation that we made at the end of the last Subsection, justified by the complicated interplay of galaxy and cluster biases modified by primordial non-Gaussianity. 

A perhaps unexpected result is that the constraints on the $f_\mathrm{NL}$ parameter in the equilateral case are almost comparable with those for the local case, and much more stringent than for the enfolded and orthogonal shapes. Since the effect of non-Gaussianity on the large-scale bias is very slight in the equilateral case one would have naively expected the opposite behavior. However, it should be recalled that the correction to the mass function (which also enters the calculations of the power spectra and the Fisher matrices) for the equilateral shape is comparable to that for the local shape. Moreover, in the former case the non-Gaussian correction to the bias is concentrated at intermediate scales, and the resulting effect on the power spectra is thus less degenerate with $\sigma_8$ with respect to the other cases. Finally, the integral over wavenumbers in the definition of the Fisher matrix weights more smaller scales. Hence a modification of  the power spectrum at a given level has a stronger impact if applied to small scales rather than large ones. In \citet*{FE09.1} it is shown how the effect of equilateral primordial non-Gaussianity on the mass function and halo bias is reduced when removing the running of $f_\mathrm{NL}$. In the next Section we show that this has the effect of loosening the constraints on the level of primordial non-Gaussianity.

\begin{table*}
\caption{The $f_{\rm NL}$--$\sigma_8$ correlation coefficients.}
\label{tab:correlation}
\begin{tabular}{lcccc}
\hline
\hline
{}&LOCAL&EQUILATERAL&ENFOLDED&ORTHOGONAL\\
\hline
\hline
$P_\mathrm{gg}$              &0.05 &0.02 &0.05 &-0.09\\
$P_\mathrm{cc}$              &0.37 &0.33 &0.27 &-0.18\\
$P_\mathrm{cg}$              &0.13 &0.13 &0.08 &-0.04\\
$P_\mathrm{cc}+P_\mathrm{cg}$       &0.13 &0.12 &0.08 &-0.05\\
$P_\mathrm{gg}+P_\mathrm{cc}+P_\mathrm{cg}$&0.04 &0.02 &0.03 &-0.04\\
\hline
\hline
\end{tabular}
\end{table*}

As a final step of our analysis, we computed the correlation coefficients between $f_\mathrm{NL}$ and $\sigma_8$ for different non-Gaussian shapes and for different combinations of LSS probes adopted. The correlation coefficient of the $i-$th and the $j-$th cosmological parameters in our set is defined as

\begin{equation}
r_{ij}\equiv \frac{\left({\mathcal F}^{-1}\right)^{ij}}
{\sqrt{\left(\mathcal F^{-1}\right)^{ii} \left(\mathcal F^{-1}\right)^{jj}}},
\label{corr-coef}
\end{equation}
where $\mathcal{F}$ is the Fisher matrix relative to the probe we are considering. For instance, the Fisher matrix for the combination of the cluster power spectrum with the cluster-galaxy cross spectrum is $\mathcal{F} = \mathcal{F}_\mathrm{cc}+\mathcal{F}_\mathrm{cg}$. According to its definition, the correlation coefficient would have value $r_{ij}=+1$ for perfectly correlated parameters, $r_{ij}=-1$ for perfectly anti-correlated ones, and $r_{ij}=0$ for uncorrelated parameters. The results are summarized in Table \ref{tab:correlation}. The first thing to note is that, while the coefficients are positive for the local, equilateral, and enfolded shapes, they are negative for the orthogonal shape. This reflects the point that, in the latter case, the confidence contours plotted in Figure \ref{fig:ellipses} are tilted in the opposite direction with respect to the former. This is a consequence of the fact that in the orthogonal case, a positive value of $f_\mathrm{NL}$ corresponds to a negative skewness of the density field and a negative correction to the linear bias, while for the other shapes considered in this work the opposite is true. A closer look to the absolute value of the correlation coefficients reveals that the constraints given by galaxy clusters alone are the most degenerate between $f_\mathrm{NL}$ and $\sigma_8$, while the constraints given by galaxies alone are almost undegenerate, in the sense that they pin down $\sigma_8$ much better than $f_\mathrm{NL}$.

The positive correlation between $\sigma_8$ and $f_\mathrm{NL}$ in terms of power spectrum of cosmic tracers can be understood by considering the effect that both parameters have on the halo bias. An increase in $\sigma_8$ has the effect of making structures less rare, and hence less biased with respect to the background matter density field. On the other hand an increase in $f_\mathrm{NL}$ in the positive direction increases the bias, except for the peculiar orthogonal case, that indeed shows an opposite degeneracy. The resulting degeneracy is however not very large because an increase in $\sigma_8$ also increases the matter power spectrum, thus partially counteracting the decrement in the halo bias.

\section{Discussion}\label{sct:discussion}

In this Section we debate some of the assumptions that have been adopted in the present paper, and discuss overlappings with other works. We begin by noting that \citet{WA10.2} forecasted the constraints on the redshift evolution of the dark energy equation of state parameter expected by the galaxy power spectrum measured by \emph{Euclid}. The error on $\sigma_8$ that they found is of the order of $\sim 7\times 10^{-2}$, thus being much larger than our result. The reasons for this mismatch are the following: \emph{i)} \citet{WA10.2} let all the cosmological parameters free to vary, while we fixed them all to their WMAP best fits, except for $f_\mathrm{NL}$ and $\sigma_8$; \emph{ii)} we used the full power spectrum information, while \citet{WA10.2} limited themselves only to the shape and position of the Baryon Acoustic Oscillation.

Recently, \citet*{SM10.1} have shown by using numerical simulations of structure formation in the presence of local non-Gaussianity, that dark matter halos tend to be on average more centrally concentrated than their counterparts in the reference Gaussian cosmology for positive $f_\mathrm{NL}$, while the opposite is true for negative $f_\mathrm{NL}$. This finding is in agreement with previous work \citep{AV03.1}, and with the naive expectation that a higher efficiency in forming high-mass objects implies. Under our assumption that the galaxy number density traces the dark matter density, this would underpin a different distribution of galaxies within dark matter halos as well. However, such an assumption is probably rough, and it is not clear if the galaxy distribution would be modified similarly to the matter distribution. Nevertheless, effects due to the inner structures of clusters show up only on very non-linear scales, that are important for cosmic shear studies, but have been neglected in the present investigation (see Section \ref{sct:results}). Thus, we stick to our choice of a standard NFW profile for the distribution of galaxies within halos for all cosmologies considered here.

\begin{figure}
\begin{center}
	\includegraphics[width=0.45\textwidth]{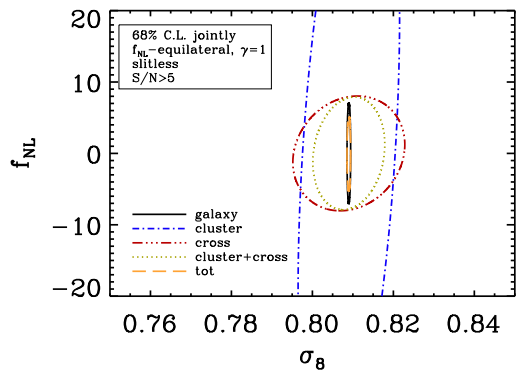} \hfill
\end{center}
\caption{The joint 68\% confidence level contours in the $f_\mathrm{NL}-\sigma_8$ plane given by different combinations of the galaxy and cluster power spectra and cluster-galaxy cross spectrum. We show only the results for the equilateral shape of the primordial bispectrum, assuming no running of the parameter $f_\mathrm{NL}$.}
\label{fig:norunning}
\end{figure}

The running of the $f_\mathrm{NL}$ parameter introduced for the equilateral shape in Eq. (\ref{eqn:eqb}) also deserves further discussion. Namely, different authors tend to use different shapes for this running, or, in some circumstances, no running at all \citep{VE09.1}. Therefore, we redid our Fisher matrix analysis by setting the running $\gamma ({\bf k}_1,{\bf k}_2,{\bf k}_3) = 1$. The results are shown in Figure \ref{fig:norunning}. As can be seen, while the constraints on $\sigma_8$ are almost unchanged those on $f_\mathrm{NL}$ are visibly loosened. For instance, the $1-\sigma$ forecasted error obtained by galaxy clusters alone goes from $\sim 13$ to $\sim 30$ when the running is removed, while that obtained by galaxies alone grows from $\sim 1.5$ to $\sim 4.6$. This kind of effect was expected, since the running we introduced had the effect of increasing the level of non-Gaussianity on sub-CMB (cluster) scales. By setting $\gamma ({\bf k}_1,{\bf k}_2,{\bf k}_3) = 1$ such an increase is removed, and the errors increase. This result stresses the importance of considering the same $f_\mathrm{NL}$ running when results from different works are compared (see also \citealt*{SH10.1}).

\begin{figure*}
\begin{center}
	\includegraphics[width=0.45\textwidth]{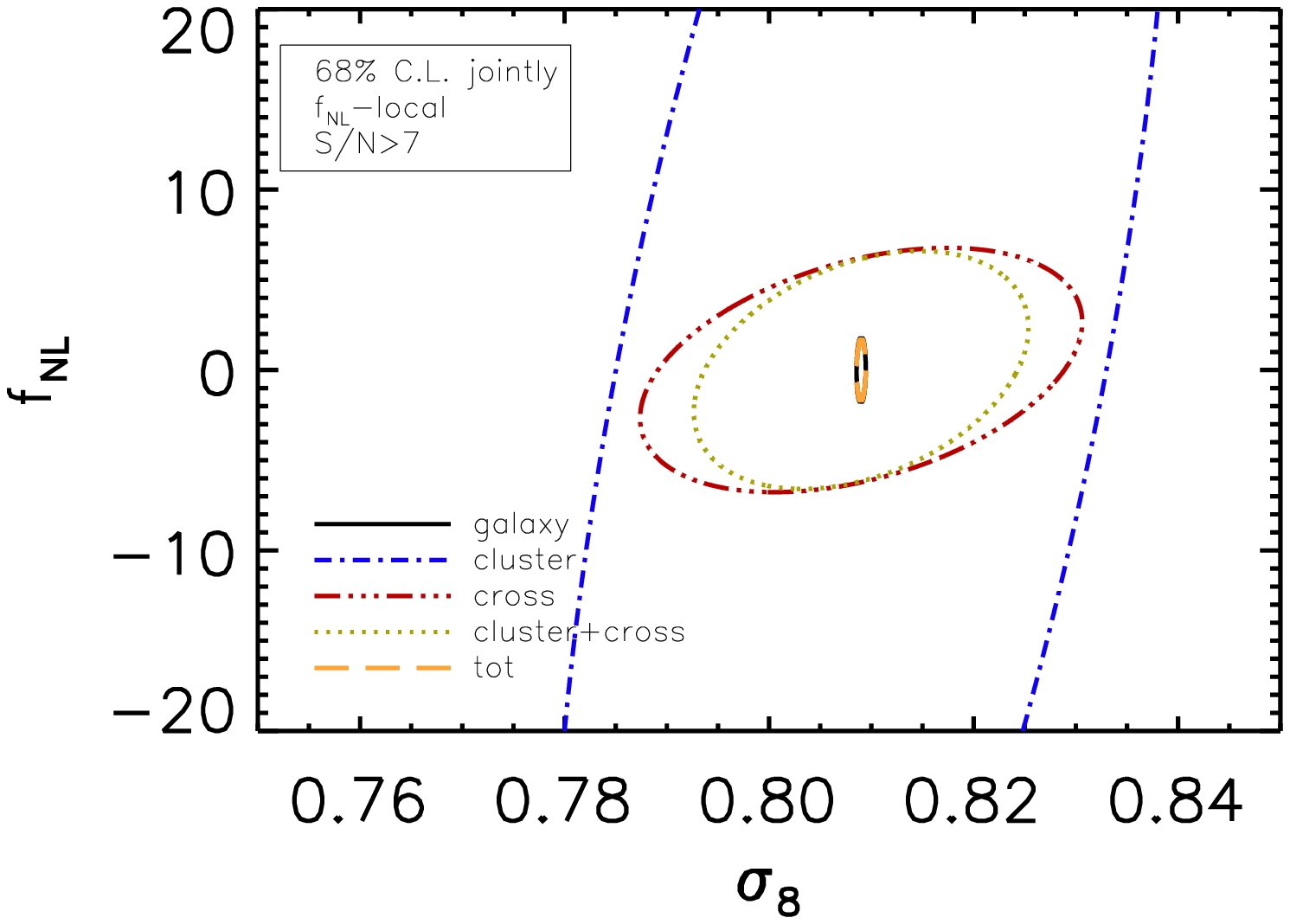}
	\includegraphics[width=0.45\textwidth]{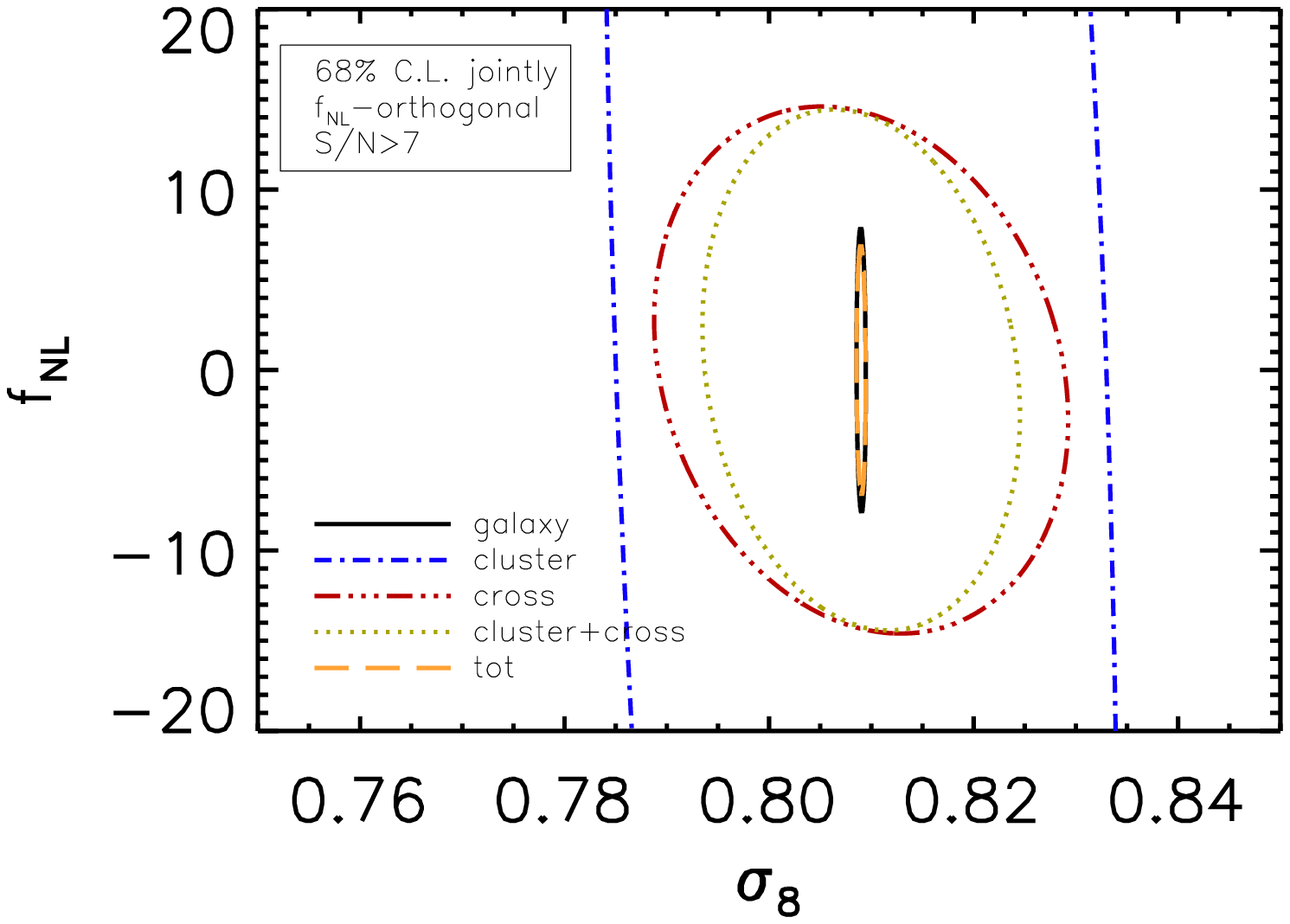}\hfill
\end{center}
\caption{The joint $68\%$ confidence level contours in the $f_\mathrm{NL}-\sigma_8$ plane given by 	different combinations of the galaxy and cluster power spectra and the cluster-galaxy cross 	spectrum, as labeled. Results are shown for the local (left panel) and orthogonal (right panel) bispectrum shapes, assuming a weak lensing cluster detection threshold of S/N $\ge 7$.}
\label{fig:ellipses_larger_sn}
\end{figure*}

In order to make our analysis more complete, we evaluated constraints on non-Gaussianity considering survey configurations different from the fiducial one described in the preceding Sections. As a first step, we changed the S/N threshold for the detection of dark matter halos in cosmic shear maps, going from S/N $\ge 5$ to S/N $\ge 7$. We again fitted the corresponding curve in Figure 1 of \citet*{BE10.2} in order to infer the minimum cluster mass to be adopted. In this case, $M_\mathrm{c}(z)$ at $z=0$ is a factor of $\sim 2$ larger than for the S/N $\ge 5$ case. Only two bispectrum shapes have been considered here: the local one, since it is the one expected to give the best constraints, and the orthogonal one, since it is the less studied one. In Figure \ref{fig:ellipses_larger_sn} we show the joint $68\%$ confidence levels on $\sigma_8$ and $f_\mathrm{NL}$ given by this new detection criterion for galaxy clusters, and in Table \ref{tab:errors_alternative} we report the numerical values of the respective errors. While the constraints given by galaxies alone are obviously unchanged, those given by clusters and the cluster-galaxy cross-correlation are significantly loosened. As a consequence, very little information is added if combining the galaxy power spectrum with power spectra involving clusters, and this remains true for the orthogonal model as well. This result implies that, while selecting cosmic shear peaks with higher S/N we are effectively selecting more massive structures, on whose statistics the effect of non-Gaussianity is larger, the respective constraining power is much too suppressed by the exponential decrement in the abundance of such objects.

\begin{figure*}
\begin{center}
	\includegraphics[width=0.45\textwidth]{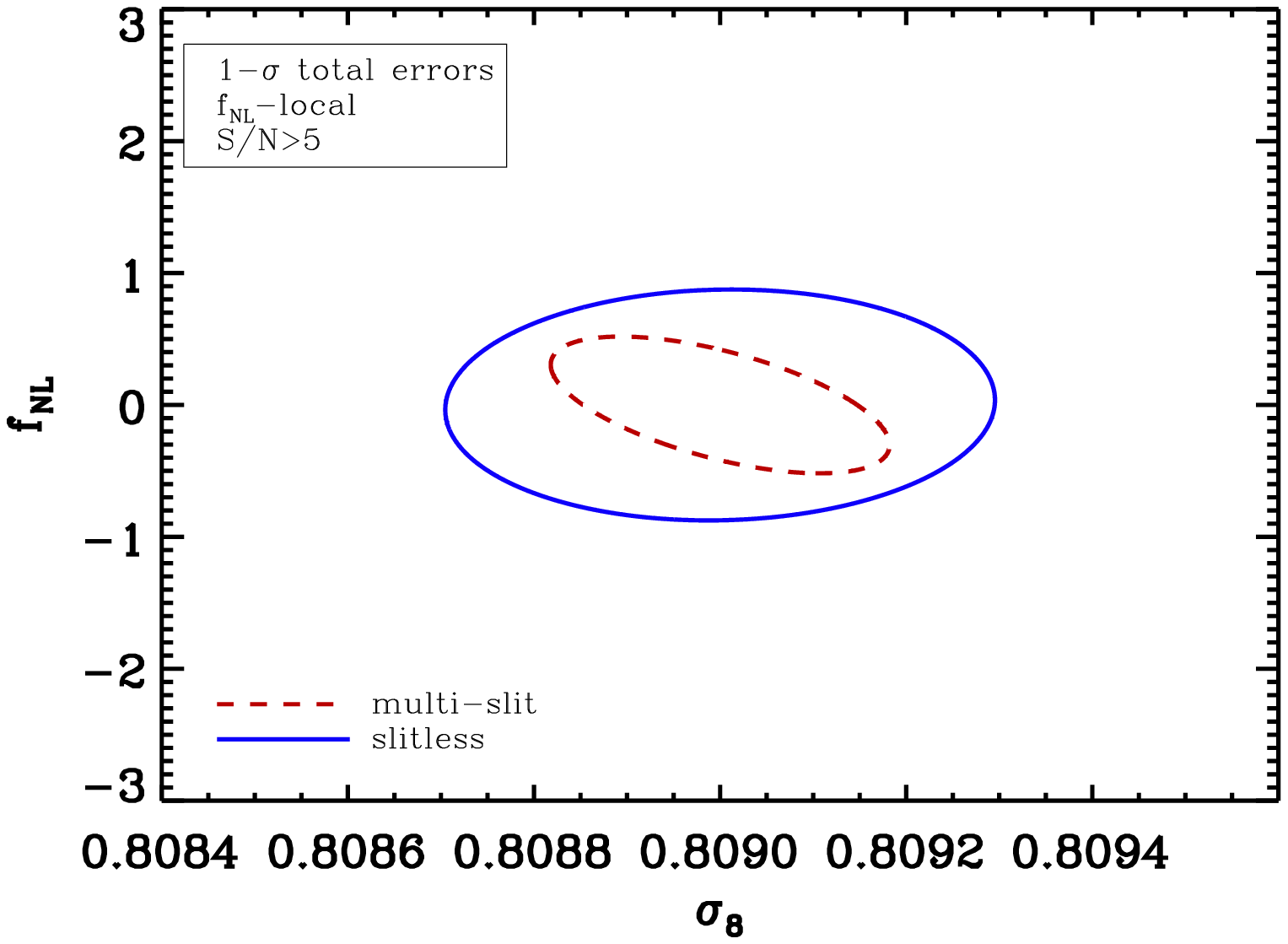}
	\includegraphics[width=0.45\textwidth]{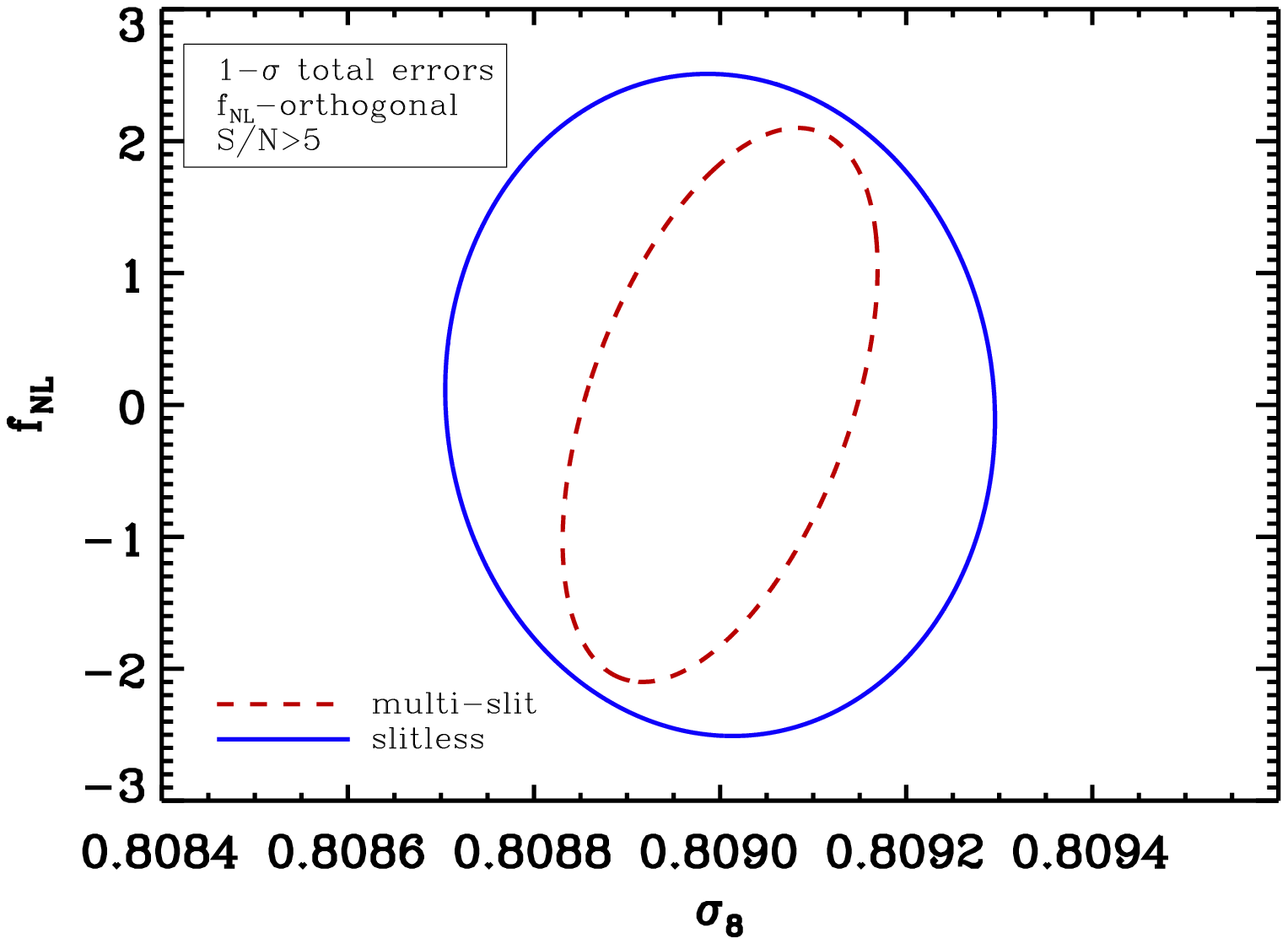}\hfill
\end{center}
\caption{The $1-$parameter $68\%$ confidence level contours in the $f_\mathrm{NL}-\sigma_8$ plane given by the combination of the galaxy power spectrum, the cluster power spectrum, and the cluster-galaxy cross spectrum. Results are shown for the local (left panel) and orthogonal (right panel) bispectrum shapes. We illustrate the comparison between the outcome for our fiducial \emph{Euclid}-like survey and for the multi-slit case discussed in the text, as labeled.}
\label{fig:ellipses_dmd}
\end{figure*}

As a second alternative survey configuration we considered a multi-slit spectroscopic instrument on the model of the Digital Micro-mirror Devices (DMDs) conceived for the SPACE mission \citep{CI09.2}. This is particularly relevant in view of future space missions alternative to \emph{Euclid}, such as WFIRST and JEDI \citep{CR05.1}. In this case selection of galaxies would occur according to their H-band flux. As a consequence Luminous Red Galaxies (LRGs) would be detected as well, thus we had to use the halo occupation distribution relative to a mixture of red and blue galaxies, rather than to blue galaxies only as we did for our fiducial survey. We found that a new constant minimum halo mass of $M_\mathrm{g}(z) = 3\times 10^{11} M_\odot h^{-1}$ reproduces fairly well the effective bias of galaxies with H-band magnitude brighter than H$_\mathrm{AB} = 22$ according to the simulations of \citet{OR10.1}. All the other parameters of the survey are left unchanged, while the redshift range of the analysis is extended to $0.1\le z\le 2.1$ (for both clusters and galaxies, although the former are unaffected by spectroscopy) in order to comply with the different observational specifications. In Table \ref{tab:errors_alternative} we report the $1-\sigma$ errors forecasted for this alternative configuration. As can be seen, in this case the constraints on both $f_\mathrm{NL}$ and $\sigma_8$ are significantly improved with respect to our fiducial \emph{Euclid}-like case. The error on the level of non-Gaussianity can for instance reach $\Delta f_\mathrm{NL}\sim 0.5$ for the local bispectrum shape. Since the cluster selection is unchanged, we interpret this improvement as due to a combination of the different galaxy selection and of the extended redshift range that is allowed by the multi-slit configuration. We expand a little bit more on this further below. This improvement is additionally emphasized in Figure \ref{fig:ellipses_dmd}, where we show the $1-$parameter 68\% confidence levels on the $\sigma_8-f_\mathrm{NL}$ plane given by the combination of the power spectra of clusters and galaxies and the cluster-galaxy cross spectrum for both the multi-slit configuration we are considering and the fiducial slitless \emph{Euclid} case. It is interesting to note that, due to the different galaxy population selected in the multi-slit case, the inclination of the confidence ellipses changes with respect to the fiducial slitless configuration.

\begin{table*}
\caption{The $1-\sigma$ forecasted errors for $f_\mathrm{NL}$ and $\sigma_8$ assuming alternative survey configurations.}
\label{tab:errors_alternative}
\begin{tabular}{lcccccccc}
\hline
\hline
{}&\multicolumn{2}{c}{LOCAL (S/N $\ge 7$)}&\multicolumn{2}{c}{ORTHOGONAL (S/N $\ge 7$)}&\multicolumn{2}{c}{LOCAL (multi-slit)}&\multicolumn{2}{c}{ORTHOGONAL (multi-slit)}\\
\hline
\hline
{} &$\Delta f_{\rm NL}$&$\Delta\sigma_8$& $\Delta f_{\rm NL}$&$\Delta\sigma_8$& $\Delta f_{\rm NL}$&$\Delta\sigma_8$& $\Delta f_{\rm NL}$&$\Delta\sigma_8$\\
\hline
$P_\mathrm{gg}$              & $1.15$&$2.96\times 10^{-4}$ & $5.15$&$2.97\times 10^{-4}$ &$0.56$&$1.90\times 10^{-4}$ &$2.69$&$1.84\times 10^{-4}$\\
$P_\mathrm{cc}$              & $37.0$&$2.00\times 10^{-2}$ & $80.5$&$1.66\times 10^{-2}$ &$6.51$&$3.25\times 10^{-3}$ &$13.8$&$2.82\times 10^{-3}$\\
$P_\mathrm{cg}$              & $4.46$&$1.43\times 10^{-2}$ & $9.63$&$1.33\times 10^{-2}$ &$1.55$&$1.93\times 10^{-3}$ &$3.54$&$1.87\times 10^{-3}$\\
$P_\mathrm{cc}+P_\mathrm{cg}$       & $4.35$&$1.08\times 10^{-2}$ & $9.52$&$1.02\times 10^{-2}$ &$1.51$&$1.60\times 10^{-3}$ &$3.43$&$1.54\times 10^{-3}$\\
$P_\mathrm{gg}+P_\mathrm{cc}+P_\mathrm{cg}$& $1.11$&$2.96\times 10^{-4}$ & $4.52$&$2.96\times 10^{-4}$ &$0.52$&$1.82\times 10^{-4}$ &$2.10$&$1.69\times 10^{-3}$\\
\hline
\hline
\end{tabular}
\end{table*}

\begin{figure*}
\begin{center}
	\includegraphics[width=0.45\textwidth]{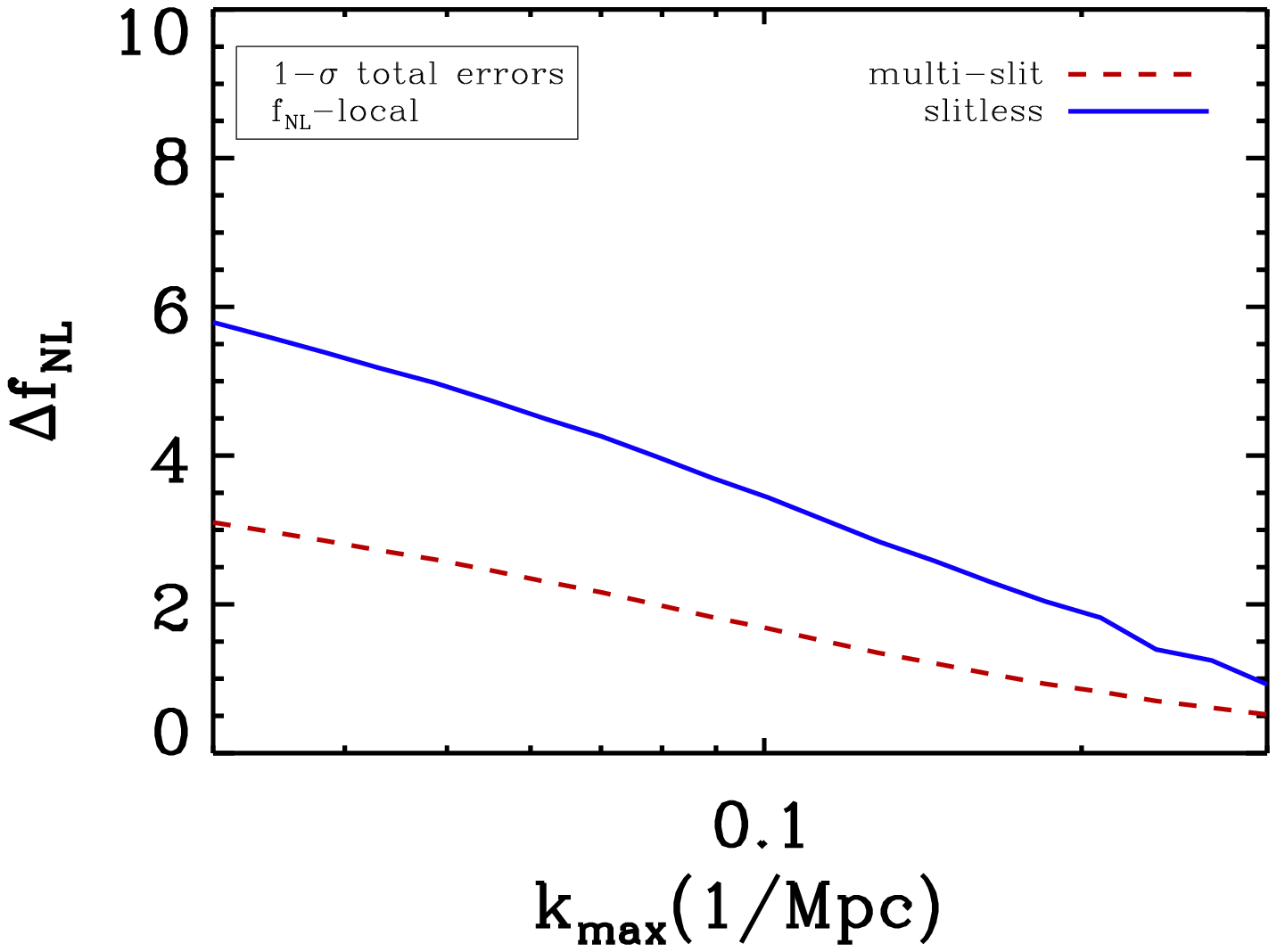}
	\includegraphics[width=0.45\textwidth]{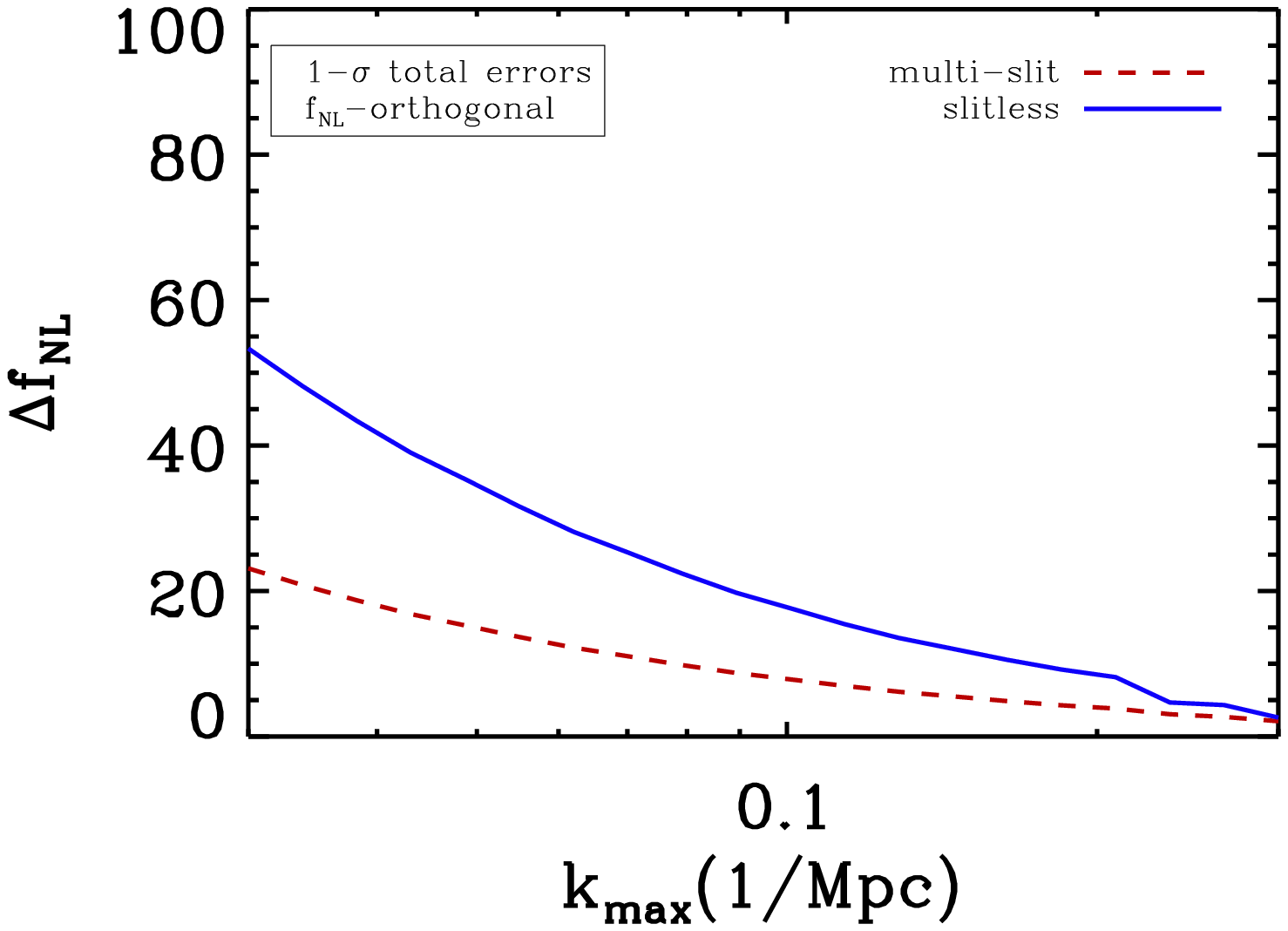}\hfill
	\includegraphics[width=0.45\textwidth]{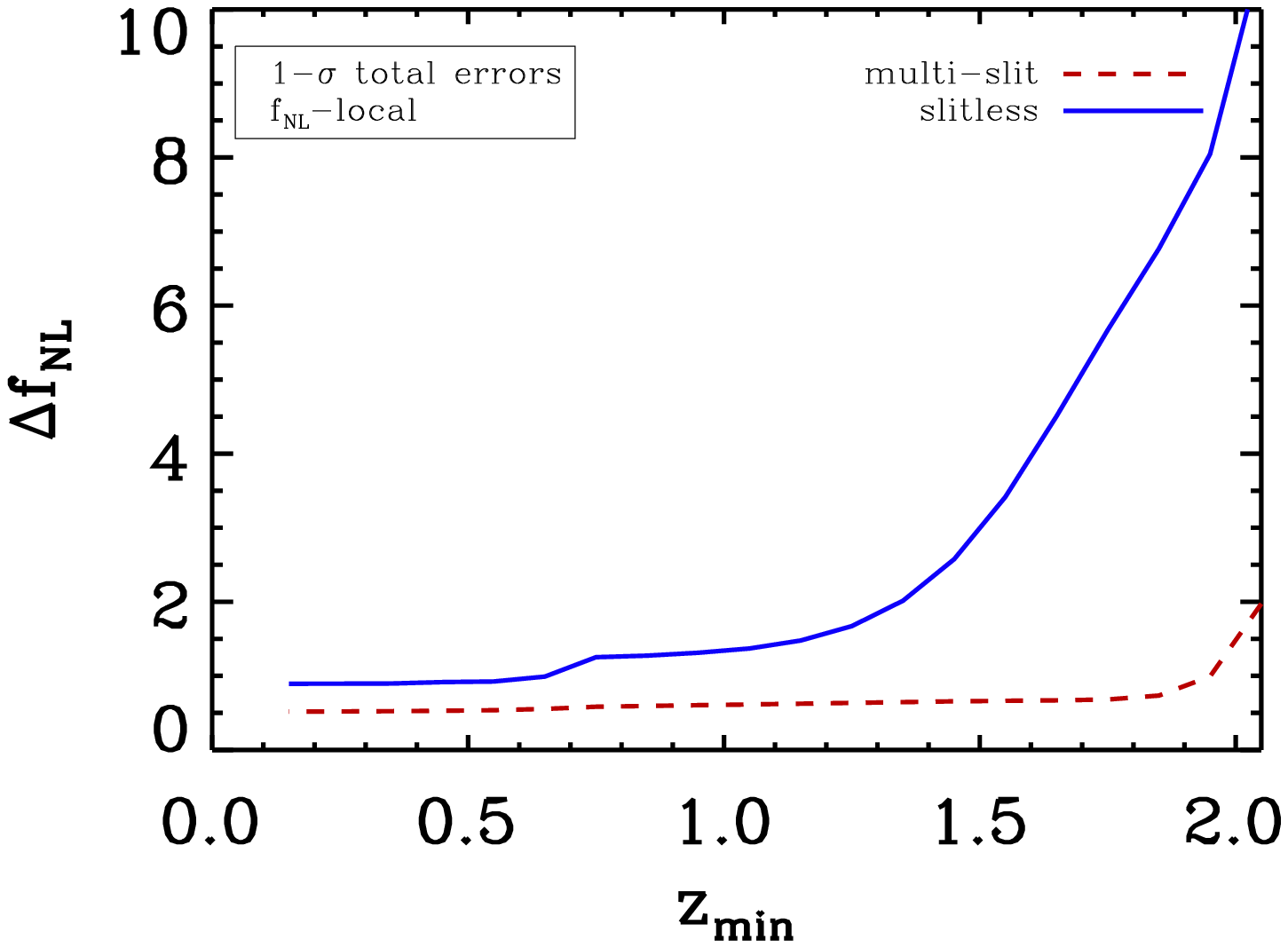}
	\includegraphics[width=0.45\textwidth]{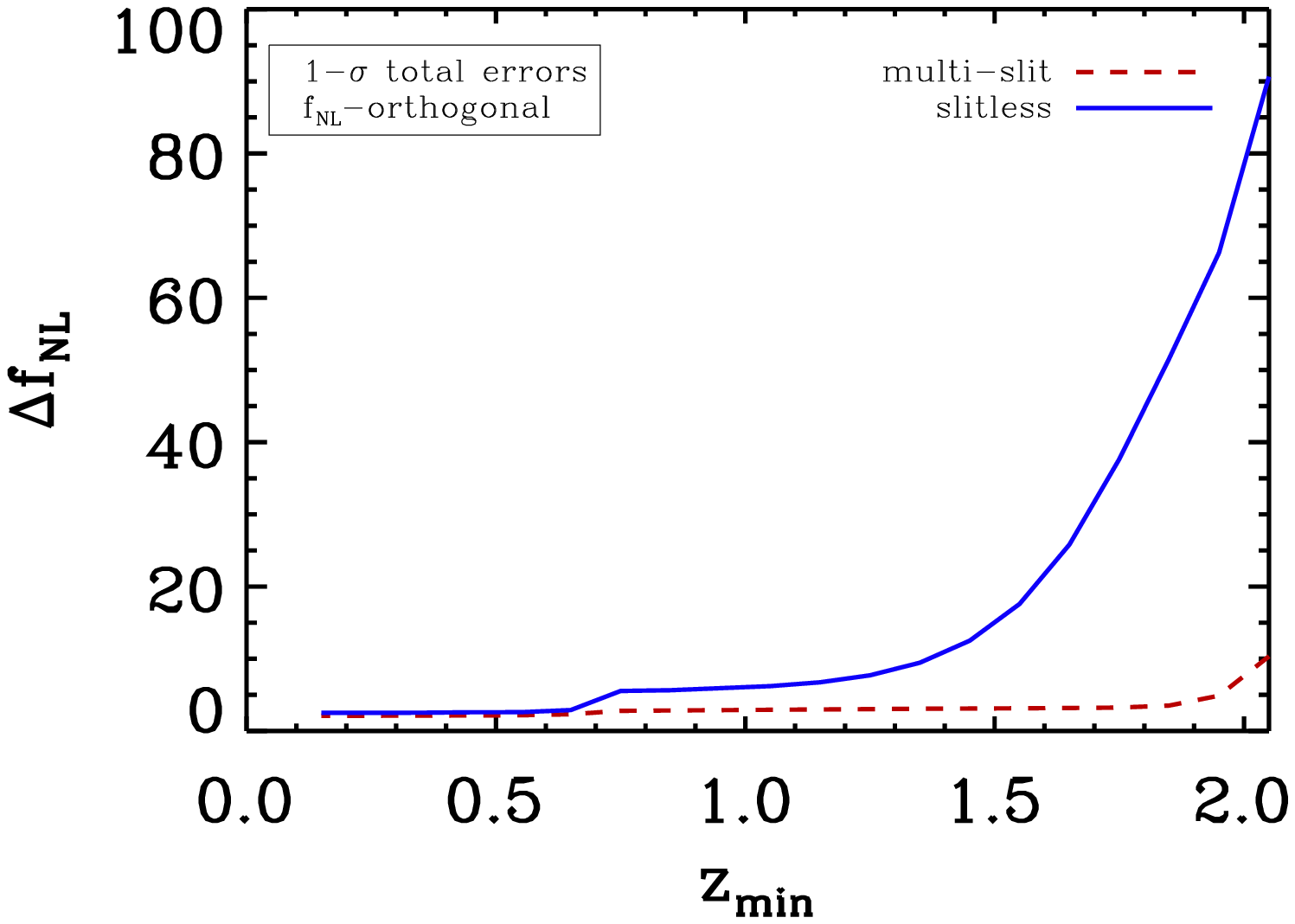}\hfill
	\includegraphics[width=0.45\textwidth]{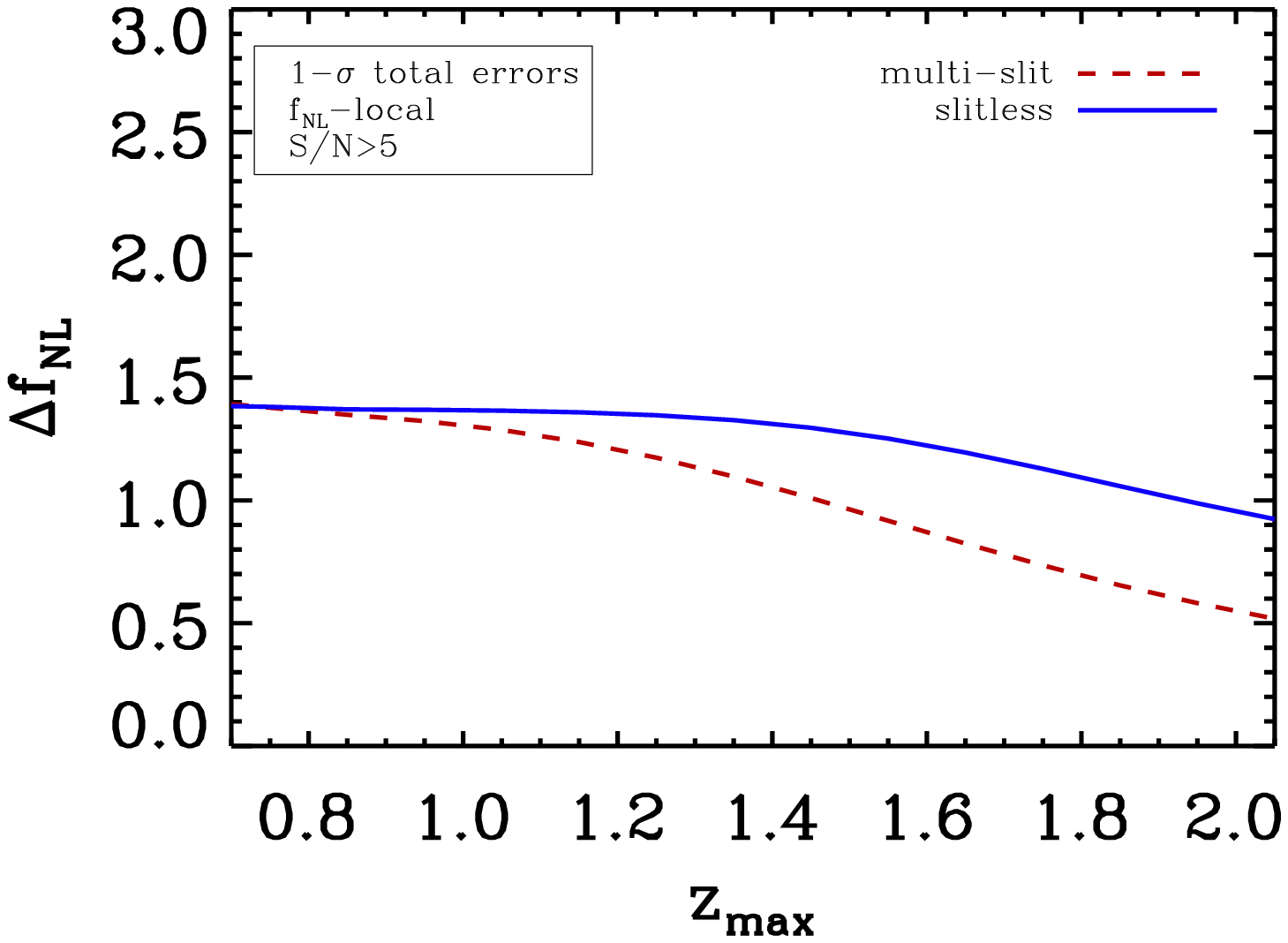}
	\includegraphics[width=0.45\textwidth]{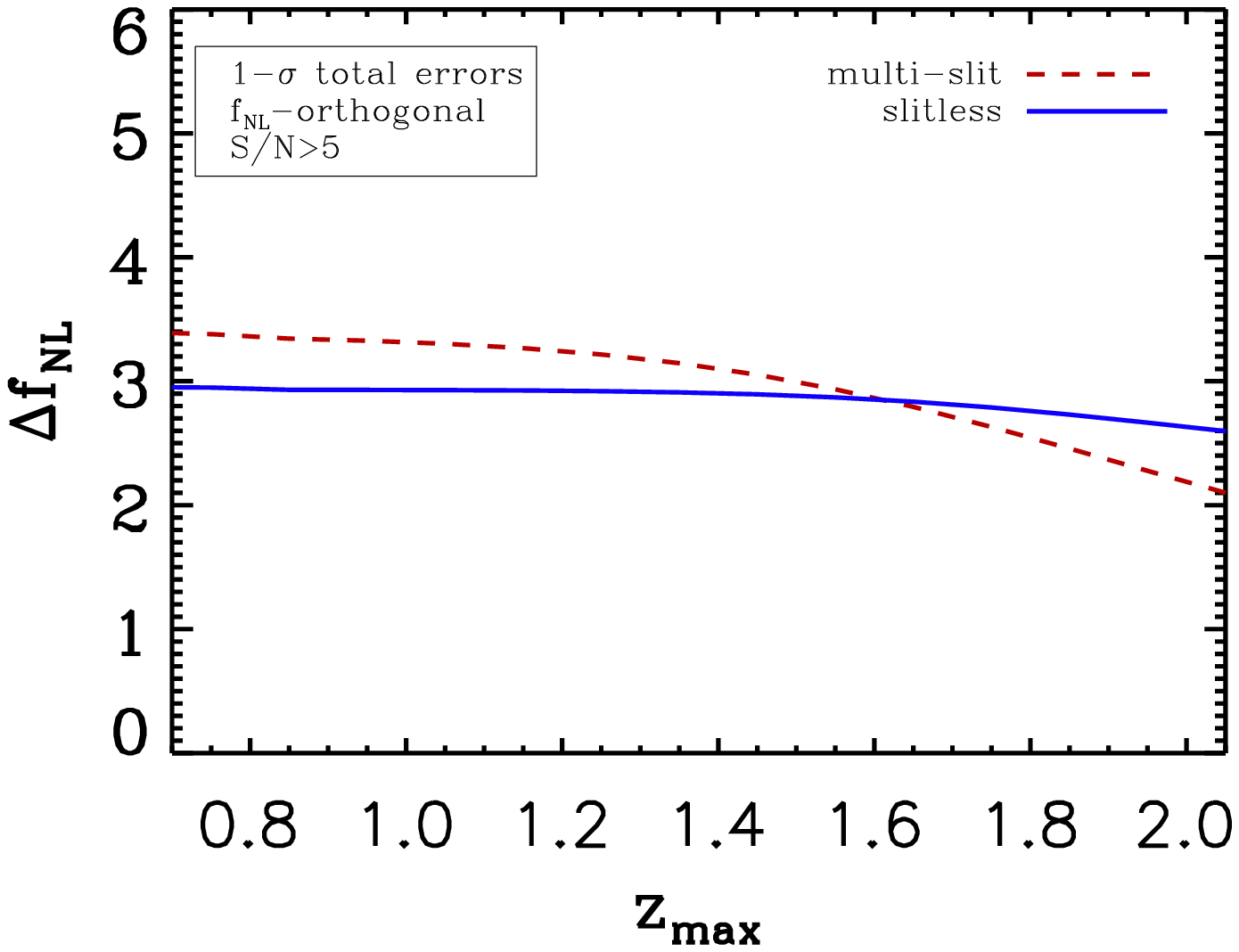}\hfill
\end{center}
\caption{The $1-\sigma$ forecasted error on the level of non-Gaussianity given by the combination of the cluster and galaxy power spectra and the cluster-galaxy cross spectrum for both the Euclid and the multi-slit survey configurations, as labeled. Results are shown for the local (left panels) and orthogonal (right panels) bispectrum shapes (please note the differences in the vertical scales). The top two panels show the trend of the error as a function of $k_\mathrm{max}$ (for fixed $0.5\le z\le 2.1$ in the slitless case and $0.1\le z\le 2.1$ in the multi-slit case), the middle two as a function of the minimum redshift (for fixed $k_\mathrm{max}=0.3$ Mpc$^{-1}$ and $z\le 2.1$), and the bottom two as a function of the maximum redshift (for fixed $k_\mathrm{max}=0.3$ Mpc$^{-1}$ and $z\ge 0.5$ for the slitless case and $z\ge 0.1$ for the multi-slit case).}
\label{fig:errors}
\end{figure*}

Finally, we estimated how the constraints on the level of non-Gaussianity changed upon modification of the redshift and scale ranges considered in the Fisher matrix analysis, for both the slitless \emph{Euclid}-like configuration and the multi-slit case. In Figure \ref{fig:errors} we show the results of changing either the minimum scale included in the analysis, $k_\mathrm{max}$, the minimum, and the maximum redshift of both galaxies and clusters. Fiducial vales are $k_\mathrm{max}=0.3$ Mpc$^{-1}$, $0.5\le z \le 2.1$ for \emph{Euclid}, and $0.1\le z \le 2.1$ for the multislit case. As one could naively expect, the error on $f_\mathrm{NL}$ decreases by increasing both the minimum scale and the maximum redshift (although in the latter case the trend is quite mild), while it increases by increasing the minimum redshift (because the maximum redshift is held fixed). As mentioned above, the differences between the slitless and the multi-slit case are due to a combination of the different selection of galaxies with the different redshift range adopted. An exception to this is given by the two middle panels of Figure \ref{fig:errors}, where the redshift ranges adopted are the same for both configurations, and thus the differences can be ascribed only to the different selection functions.

\section{Summary and conclusions}\label{sct:conclusions}

We adopted the physically motivated halo model in order to compute the effect of different kinds of primordial non-Gaussianity on the power spectrum of galaxies and galaxy clusters, as well as on the cluster-galaxy cross spectrum. Specifically, we considered galaxies selected spectroscopically according to their H$\alpha$ flux, and galaxy clusters selected as the largest S/N peaks in cosmic shear maps, having in mind future wide field optical/near-infrared surveys such as \emph{Euclid} and WFIRST. We additionally performed a Fisher matrix analysis in order to forecast the errors on the joint determination of the level of non-Gaussianity $f_\mathrm{NL}$ and the amplitude of the matter power spectrum $\sigma_8$. The main findings of this work can be summarized as follows.
\begin{itemize}
\item The non-Gaussian corrections to the power spectrum of tracers of the LSS resembles the modifications to the large-scale bias of dark matter halos, as one might naively expect. The largest effect is seen for the local bispectrum shape, while the smallest appears for the enfolded shape. The power spectrum of massive galaxy clusters is more heavily modified than the spectrum of galaxies, because the former are originally more biased than the latter. The modification to the cluster-galaxy cross spectrum lies somewhere in between.
\item Galaxies have a much higher constraining power on both $f_\mathrm{NL}$ and $\sigma_8$ as compared to galaxy clusters, due to the much lower abundance of the latter that is not adequately compensated by the larger effect on the relative power spectrum. Assuming a \emph{Euclid}-like survey, while spectroscopically selected galaxies can constrain $f_\mathrm{NL}$ to the level of a few and $\sigma_8$ to the level of $\sim 3 \times 10^{-4}$, errors on parameters derived by clusters alone are at the level of $\Delta f_\mathrm{NL} \sim 10$ and $\Delta\sigma_8 \sim 8\times 10^{-3}$ (with some dependence on the primordial bispectrum shape).
\item When considering the cluster-galaxy cross spectrum alone, the forecasted constraints on $\sigma_8$ are comparable to the cluster-only constraints, while the constraints on $\sigma_8$ are improved by more than one order of magnitude, reaching a predicted error comparable with that coming from galaxies alone, $\Delta f_\mathrm{NL} \sim$ a few. This result highlights the high complementarity of the cluster power spectrum and the cluster-galaxy cross spectrum as cosmological probes.
\item While the constraints on $\sigma_8$ coming from the galaxy power spectrum alone are almost unchanged when combined with the cluster-galaxy cross correlation, the constraints on $f_\mathrm{NL}$ are improved. This is true only to a slight level for the local and equilateral bispectrum shapes, however the error on $f_\mathrm{NL}$ can be reduced by a factor of $\sim 2$ for the enfolded and orthogonal cases. The addition of the cluster power spectrum carries only slight change to this conclusion.
\item As expected the parameters $f_\mathrm{NL}$ and $\sigma_8$ are degenerate with respect to the power spectra of LSS tracers. This degeneracy is more marked for the cluster power spectrum, with correlation coefficients reaching up to $0.3-0.4$. The degeneracy on the other hand is almost insignificant for the galaxy power spectrum, in the sense that the constraining power weights much more on $\sigma_8$ than on $f_\mathrm{NL}$. The degeneracy with respect to the cluster-galaxy cross spectrum lies in between the former two, with the exception of the orthogonal bispectrum shape.
\item We considered several survey configurations alternative to the fiducial \emph{Euclid}-like one, finding that a multi-slit spectroscopic instrument would allow more stringent constraints both on $f_\mathrm{NL}$ and $\sigma_8$. This improvement is due to a combination of the fact that with this other arrangement the selected galaxy population is different, and that the data analysis can be pushed down to a lower minimum redshift. This result is interesting with respect to future wide field survey concepts alternative to \emph{Euclid}, such as WFIRST. 
\end{itemize}

The improvement in the constraints on both $f_\mathrm{NL}$ and $\sigma_8$ when combining the cluster and/or galaxy power spectra with the cluster-galaxy cross spectrum is reminescent of the fact that the latter is sensitive to different scales in a different way with respect to the former. The fact that this improvement is more important for the enfolded and orthogonal bispectrum shapes is extremely interesting, since these models are still relatively poorly studied compared to the equilateral and, almost ubiquitous, local shapes. Additionally, the amazing constraining power of the galaxy power spectrum, even when considered alone, stresses the importance of the spectroscopy part for future \emph{Euclid}-like missions when it comes to restrict cosmology on the basis of the spatial distribution of objects.

The present work reinforces the conclusion that the spatial distribution of tracers of the LSS, especially galaxies, is an incredibly powerful probe for primordial non-Gaussianity, thanks to the very strong impact that the latter has on the large scale bias of dark matter halos \citep*{CA08.1,CA10.1}. The combination of cluster and/or galaxy power spectra with the cross spectrum of clusters and galaxies significantly improves the forecasted constraints, especially for the least studied non-Gaussian shapes. Merging all the information from future wide field surveys such as \emph{Euclid} and WFIRST promise to bring constraints on $f_\mathrm{NL}$ to the unity level, and constraints on $\sigma_8$ to the level of $\sim 10^{-4}$, in both cases superior to future CMB experiments \citep{SE09.1,VE09.1}.

\section*{Acknowledgments}
We acknowledge financial contributions from contracts ASI-INAF I/023/05/0, ASI-INAF I/088/06/0, ASI I/016/07/0 'COFIS', ASI '\emph{Euclid}-DUNE' I/064/08/0, ASI-Uni Bologna-Astronomy
Dept. '\emph{Euclid}-NIS' I/039/10/0, and PRIN MIUR 'Dark energy
and cosmology with large galaxy surveys'. We are grateful to N. Bartolo, A. Gonzalez and B. Sartoris for stimulating discussions about the manuscript, and to A. Orsi for sharing the results of his simulations. We wish to thank the anonymous referee for useful comments that allowed us to improve the presentation of our work.

{\small
\bibliographystyle{aa}
\bibliography{./master}
}

\end{document}